\documentclass[a4paper,11pt]{article}
\pdfoutput=1
\usepackage{amssymb,amsmath,amsfonts,makeidx,placeins,pbox,multirow}
\usepackage{graphicx,rotate,subcaption,color,slashed,cite,caption,epstopdf,verbatim}
\usepackage{longtable,tabu,booktabs,cancel}
\usepackage{array}
\usepackage{pstricks}
\usepackage{color}
\usepackage{amsmath}
\usepackage{mathtools}
\usepackage{amssymb}
\usepackage{bbold}
\usepackage[normalem]{ulem}
\usepackage{caption}
\usepackage{subcaption}
\usepackage{dsfont}

\newcolumntype{P}[1]{>{\centering\arraybackslash}p{#1}}
\usepackage[colorlinks=true,
            linkcolor=magenta,
            urlcolor=blue,
            citecolor=blue]{hyperref}
\evensidemargin=\oddsidemargin
\newcommand{\beq}{\begin{equation}}
\newcommand{\eeq}{\end{equation}}
\newcommand{\bea}{\begin{eqnarray}}
\newcommand{\eea}{\end{eqnarray}}

\def\bar {\overline}

\def\to {\rightarrow}

\def\bea {\begin{eqnarray}}
\def\eea {\end{eqnarray}}
\def\n {\nonumber}

\def\barr{\begin{array}}
\def\earr{\end{array}}
\def\to{\end{rightarrow}}

\def\gev{\ensuremath{\mathrm{Ge\kern -0.1em V}}}

\usepackage{color}
\definecolor{cornellred}{rgb}{0.7, 0.11, 0.11}

\def\hlvs#1{\textcolor{magenta}{#1}}
\usepackage{tikz}
\usetikzlibrary{positioning,arrows}
\usetikzlibrary{decorations.pathmorphing}
\usetikzlibrary{decorations.markings}

\begin{document}

\tikzset{
vector/.style={decorate, decoration={snake,amplitude=.6mm}, draw=red},
scalar/.style={dashed, draw=blue},
fermion/.style={draw=black, postaction={decorate},
        decoration={markings,mark=at position .55 with {\arrow[draw=black]{>}}}}
}
\begin{center}
{\Large \bf Search for Quadruplet Scalars using Boosted Decision Trees at the LHC \\
\vspace{0.3cm}
} 
\vspace*{0.4cm} {\sf $^{a}$Amit Chakraborty\footnote{amit.c@srmap.edu.in}, $^{a}$Shreecheta Chowdhury\footnote{shreecheta\_c@srmap.edu.in}, $^{b}$Nilanjana Kumar\footnote{nilanjana.kumar@gmail.com}, $^{c}$Vandana Sahdev \footnote{vandanasahdev20@gmail.com}} \\
\vspace{6pt} {\small } {\em  $^{a}$ Department of Physics, SRM University-AP, Amaravati, Mangalagiri 522240, India\\$^{b}$Department of Physics, Chettinad Institute of Technology, Chettinad Academy of Research and Education, Chennai, Tamilnadu 603103, India,\\
$^{c}$Department of Physics and Astrophysics, University of Delhi, Delhi 110007, India} \\
\normalsize
\end{center}
%==========================================================================================================
\vspace{-0.4cm}
%===========================================================================================================
\begin{abstract}
Beyond the Standard Model scenarios introduce additional scalar and fermion multiplets, which influence neutrino mass generation mechanisms and yield distinctive collider signatures. This work focuses on a particular scenario involving a fermion quintuplet and a scalar quadruplet. The study examines the production and decay of the scalar quadruplet components at the Large Hadron Collider (LHC), emphasizing how their decay patterns, {\em fermiophobic} versus {\em fermiophilic}, depend on mass differences and Yukawa couplings with the fermion multiplets. This study provides an overview of possible signals at the LHC, along with a detailed collider analysis focused on final states containing at least four leptons and two jets, in which the masses of the scalars and fermions are reconstructed successfully. Standard Model backgrounds are also incorporated in the study, with multivariate techniques leveraged via Boosted Decision Trees. Results indicate discovery potential for scalar masses around $600–700$ GeV and exclusion sensitivity extending beyond $1$ TeV, highlighting the promising experimental signatures of the model and its role in probing new physics at colliders.
\end{abstract}
%==========================================================================================================
\vspace{-0.4cm}
%===========================================================================================================
\newpage
\tableofcontents

%----------------------------------------------------------------------------------------------------------
\section{Introduction}
%----------------------------------------------------------------------------------------------------------
\label{sec:intro}

The Standard Model (SM), while spectacularly successful, is fundamentally incomplete. Its inability to explain phenomena such as neutrino masses, the nature of dark matter, and the hierarchy problem, among other issues, strongly motivates physics beyond the SM (BSM). Many sophisticated models are proposed to overcome the shortcomings of SM 
\cite{Haber:1984rc,Maldacena:1997re,Gubser:1998bc,Witten:1998qj,Randall:1999ee,Georgi:1974sy,Pati:1974yy,Arkani-Hamed:1998jmv,Kaplan:1983fs,Agashe:2004rs,Mohapatra:1974gc,Senjanovic:1975rk,Chacko:2005pe}.
In many of these models, extra scalar degrees of freedom are either necessary or appear as a consequence of the fundamental theory. 
However, simple extensions of SM such as Two Higgs Doublet Model (2HDM) \cite{Branco:2011iw}, Type-II seesaw with scalar triplets \cite{Schechter:1980gr}, 
appear with additional scalar multiplets. In these models, tiny neutrino masses arise via the seesaw mechanism. However, tree-level seesaw models face tension between the seesaw scale and couplings \cite{Boucenna:2014zba,Ding:2014nga}. 
A resolution to this is to suppress the tree-level contribution and make the leading contribution come from loop-level diagrams of the dimension-5 operator \cite{Bonnet:2012kz,Ma:2006km}, or to raise
the dimension of the effective operator\cite{Babu:2009aq,Kumericki:2012bh,Nomura:2017abu,Ghosh:2018drw}. The appearance of large scalar multiplets is very natural in these types of models\cite{Giarnetti:2023dcr,Giarnetti:2023osf}.
Another alternative method to lower the seesaw scale is by the ``cascade seesaw mechanism''\cite{Ding:2014nga}, where the neutrino masses are generated at tree level via a combination of Type-I and Type-III seesaw mechanisms. These models include both scalar and fermion multiplets as part of their exotic particle content.

The presence of one or more scalar and/or fermion multiplets in models such as GUT \cite{Georgi:1974sy}, Left-Right Symmetric model\cite{Senjanovic:1975rk,KumarAgarwalla:2018nrn}, Little Higgs Models \cite{Low:2002ws}, Composite Higgs Models \cite{Vignaroli:2012sf}, R$\nu$MDM \cite{Cai:2016jrl} opens up the possibility of unique collider signatures. 
For instance, Ref.~\cite{Dubey:2025sfh} demonstrates how leptoquarks enhance vector-like lepton production, yielding distinctive signatures. Ref.~\cite{Ashanujjaman:2022cso} introduces two leptonic multiplets under different SU(2) representations to address SM shortcomings. In Composite Higgs Models \cite{Kumar:2025rqs,Banerjee:2022xmu}, pNGB scalars may decay via heavy fermionic resonances, which results in unique collider signatures.
In this work, we explore a model featuring a fermion quintuplet (with zero hypercharge) and a scalar quadruplet (with hypercharge $1/2$), which together can generate neutrino masses through  tree-level as well as loop-level mechanisms \cite{Ma:2014zda,Nomura:2017abu,Kumar:2019tat}. 

We explore the production and decays of 
the quadruplet scalar (comprising neutral, singly charged, and doubly charged components) at the LHC. 
The decay modes of these quadruplet scalars depend critically on the mass difference between the scalar and fermion multiplets. The Yukawa coupling between the scalar and fermion multiplets also plays a major role in their decays. We show that there are two distinct decay patterns where the scalar can either decay to the gauge bosons
directly (when the Yukawa coupling is small) or they first decay to quintuplet fermions which subsequently decay to SM particles (when the Yukawa coupling is large). These two scenarios are named {\em fermiophobic} and {\em fermiophilic}, respectively. 
We discuss the possible collider signatures in these two scenarios, along with the existing limits from the collider studies.

The introduction of two distinct $SU(2)$ multiplets as the BSM particle content poses significant challenges for the phenomenology. 
We discuss the pair and associated production of charged scalars within the mass range $300-1000$ GeV, at 14 TeV LHC. Beyond identifying channels with superior discovery potential, this paper focuses on those enabling mass reconstruction of exotic scalars and fermions—crucial for new particle discovery.
Our detailed analysis of signal processes reveals that the final state with 4 or more leptons and 2 jets ({\em fermiophilic} scenario) offers sufficient sensitivity to probe both the quartet scalar and quintuplet fermions at the LHC. 
We analyze the most relevant SM background that exhibits overlapping kinematics, with a focus on the key reconstructed observables that offer some discrimination power. The sensitivity in these channels is estimated by performing a multivariate analysis (MVA) using the Boosted Decision Tree (BDT) framework. We also demonstrate the reconstruction of the quintuplet scalars and quintuplet fermion from their decay products.

%%%%%%%%%%%%%%%%%%%%%%%%%%%%%%%%%%%%%%%%%%%%%%%%%%%%%%%%%%%%%%%%%%%%%%%%%%
\section{Model}
\label{sec:model}
%%%%%%%%%%%%%%%%%%%%%%%%%%%%%%%%%%%%%%%%%%%%%%%%%%%%%%%%%%%%%%%%%%%%%%%%%%%%
This work is motivated by the models studied previously in \cite{Kumericki:2012bh,Nomura:2017abu,Kumar:2019tat}. 
To keep the analysis minimal and transparent, 
we consider the simplest scenario where the BSM particle content comprises a quadruplet scalar ($\Phi_4=\left( \phi^{++}, \phi^{+}_2, \phi^{0}, \phi^{-}_1 \right)^T$) and a quintuplet fermion ($\Sigma_R =\left[\Sigma_1^{++},\Sigma_1^{+},{\Sigma^0}, \Sigma_{2}^-, \Sigma_{2}^{--} \right]_R^T$) only.
We denote the scalar and the fermion masses by $M_{\Phi}$ and $M_{\Sigma}$, respectively
\footnote{The mass of the scalar comes from the scalar potential
$V=\mu^2 H^{\dagger} H + M_{\Phi}^2 \Phi_{4}^{\dagger} \Phi_4$.}. 

The kinetic term of the quadruplet scalar, obtained by expanding the tensors, is
%%%%%%%%%%%%%%%%%%%%%%%%%%%%%%
\begin{align}
|D_\mu \Phi_4|^2 = \sum_{m= -\frac32,-\frac12,\frac12,\frac32} \biggl| & \left[ \partial_\mu - i \left(\frac12+m \right) e A_\mu - i \frac{g}{c_W} \left(m - \left( \frac12+m \right) s_W^2 \right) Z_\mu \right] (\Phi_4)_{m} \nonumber \\
& + \frac{ig}{\sqrt{2}} \sqrt{ \left(\frac32 + m \right) \left(\frac52 -m \right) } W^+_\mu (\Phi_4)_{m-1} \nonumber \\
& + \frac{ig}{\sqrt{2}} \sqrt{ \left(\frac32 - m \right) \left(\frac52 +m \right) } W^-_\mu (\Phi_4)_{m+1} \biggr|^2,
\end{align}
%%%%%%%%%%%%%%%%%%%%%%%%%%%%%%
where $m$ is the weak isospin and $s_W~(c_W) = \sin \theta_W ~(\cos \theta_W)$, $\theta_W$ being the Weinberg angle.
From here, we obtain the interactions responsible for the production of the scalars viz.,
%%%%%%%%%%%%%%%%%%%%%%%%%%%%%%
\begin{align}
|D_\mu \Phi_4|^2 \supset &- i e A_\mu \phi_1^+ \partial^\mu \phi_1^- - \frac{i g}{c_W} \left( \frac{3}{2} - s_W^2 \right) Z_\mu \phi_1^+ \partial^\mu \phi_1^- \nonumber \\
&- i e A_\mu \phi_2^+ \partial^\mu \phi_2^- - \frac{i g}{c_W} \left( \frac{1}{2} - s_W^2 \right) Z_\mu \phi_2^+ \partial^\mu \phi_2^-  \nonumber \\
& - 2 i e A_\mu \phi^{++} \partial^\mu \phi^{--} - \frac{i g}{c_W} \left( \frac{3}{2} - {2} ~s_W^2 \right) Z_\mu \phi^{++} \partial^\mu \phi^{--} \nonumber \\
& - i \sqrt{\frac{3}{2}} g \left(W^+_\mu \phi^{--}  \partial^\mu \phi_2^+ + W^-_\mu \phi_2^- \partial^\mu \phi^{++}\right) + ~h.c.
\label{eq:prod_gauge}
\end{align}
%%%%%%%%%%%%%%%%%%%%%%%%%%%%%%
Here, the neutral component of ${\Phi}_4$ gets a non-zero VEV, $\frac{v_4}{\sqrt{2}}$, induced by virtue of its quartic coupling with the SM doublet Higgs. The VEV is constrained from the measurement of the $\rho$ parameter \cite{Kumericki:2012bh,Yu:2015pwa,Nomura:2017abu}.% CHEKKKKKKKK. 
\footnote{These models are further constrained from muon $(g-2)$ and lepton flavor violation experiments which is discussed in \cite{Nomura:2017abu,Kumar:2019tat}.}.
The decay of the scalars to the gauge bosons is governed by the following terms:
%%%%%%%%%%%%%%%%%%%%%%%%%%%%%%
\begin{align}
|D_\mu \Phi_4|^2 \supset &\frac{\sqrt{3}}{2} \frac{g^2}{cw} v_4 \left(2 - s_W^2 \right)W^+_\mu Z^\mu \phi_1^- + \frac{g^2}{cw} v_4 s_W^2 W^+_\mu Z^\mu \phi_2^- + \sqrt{\frac{3}{2}} {g^2} v_4 W^+_\mu W^{+\mu} \phi^{--} + ~h.c.
\label{eq:decay_gauge}
\end{align}
%%%%%%%%%%%%%%%%%%%%%%%%%%%%%%
It is important to note that the decay of the $\Phi_4$ to gauge bosons depends upon its VEV, $v_4$. 

The Yukawa interaction between the quadruplet scalar and the quintuplet fermion can be expressed as
%%%%%%%%%%%%%%%%%%%%%%%%%%%%%%
\begin{align}
-{\cal L}_{yuk}& \supset (y_{\nu})_{ij} [\bar L_{L_i} \tilde\Phi_4 \Sigma_{R_j} ]+{\rm h.c.} \n \\
&= (y_\nu)_{ij}
\left[
\bar\nu_{L_i} \left(\Sigma^{++}_{1R_j} \phi^{--} + \frac{\sqrt3}{2} \Sigma^+_{1R_j} \phi_1^-  + \frac{1}{\sqrt2}\Sigma^0_{R_j} \phi^{0*} + \frac{1}{2}\Sigma^-_{2R_j} \phi_{2}^+ \right)\right.
\n\\
&\left. +
\bar\ell_{L_i} \left( \frac12 \Sigma^+_{1R_j} \phi^{--} + \frac{1}{\sqrt2}\Sigma^0_{R_j} \phi_1^- + \frac{\sqrt3}{2}\Sigma^-_{2R_j} \phi^{0*} + \Sigma^{--} _{2R_j}\phi_{2}^{+}  \right)\right] +{\rm h.c.}
\label{eq:yukawa}
\end{align}
%%%%%%%%%%%%%%%%%%%%%%%%%%%%%%
This Yukawa interaction induces diagrams that lead to tiny neutrino masses \cite{Nomura:2017abu,Kumar:2019tat}. 
The induced VEV allows decent Yukawa coupling for %$\cal{O}$(1 TeV) 
quintuplet masses $\leq 1$ TeV as neutrino mass, $m_\nu \propto v_4^2 ~y_\nu M_\Sigma^{-1}  y_\nu^T$ \cite{Nomura:2017abu}. For 
example, we find that for small $y_\nu$ ($\sim0.0007$), $v_4\sim$ 1 GeV and for large $y_\nu$ ($\sim 1$), $v_4\sim0.0002$ GeV.  In order to generate non-zero masses for all three SM neutrinos, we need three copies of the quintuplets \cite{Casas:2001sr,Sahdev:2024wie}. For the purpose of this study, we assume that $M(\Sigma_2, ~\Sigma_3) > > M(\Sigma_1)$ and include only one copy of the fermion multiplet $\Sigma_R$.

If the fermions are heavier than the scalars ($M_\Sigma > M_\Phi$), the scalars can only decay to the gauge bosons, which follows from Eqs.~\ref{eq:decay_gauge} and \ref{eq:yukawa}\footnote{For a detailed study of the case $M_\Sigma > M_\Phi$, please see \cite{Kumar:2021umc}}. In the converse case, the scalars can decay to the fermions as well, in addition to the gauge bosons. Thus, in the latter case, we observe two scenarios: 
%%%%%%%%%%%%%%%%%%%%%%%%%%%%%%
\begin {itemize}
\item Fermiophobic: When $y_\nu$ is small.
\item Fermiophilic: When $y_\nu$ is large.
\end{itemize}
%%%%%%%%%%%%%%%%%%%%%%%%%%%%%%

Since we consider $M_\Phi > M_\Sigma$, the quintuplets do not decay to scalars. 
However, by virtue of their mixing with the SM leptons, which arises due to the Yukawa interaction in Eq.~\ref{eq:yukawa} when $\phi^0$ obtains the VEV, they can decay directly to an SM fermion and a gauge boson.
The mixing of neutral fermions with SM fermions can be written as
%%%%%%%%%%%%%%%%%%%%%%%%%%%%%%
\beq
\mathcal{L}_{mass,N}  \supset - \frac{1}{2} {\begin{pmatrix} \bar{\nu}_L & \bar{(\Sigma_R^0)^C} \end{pmatrix}}  {\begin{pmatrix} 0 & \frac{1}{\sqrt{2}} y_{\nu} \frac{v_4}{\sqrt{2}} \\ \frac{1}{\sqrt{2}} y_{\nu}^T \frac{v_4}{\sqrt{2}}  & M_{\Sigma} \end{pmatrix}}   {\begin{pmatrix} (\nu_L)^C \\ \Sigma_R^0 \end{pmatrix}} + h.c.
\eeq
%%%%%%%%%%%%%%%%%%%%%%%%%%%%%%
Similarly, the mixing of charged fermions with SM fermions can be written as
%%%%%%%%%%%%%%%%%%%%%%%%%%%%%%
\beq
\mathcal{L}_{mass,C}  \supset - {\begin{pmatrix} \bar{l}_L & \bar{(\Sigma_R^+)^C} \end{pmatrix}}  {\begin{pmatrix} y_l \frac{v}{\sqrt{2}} & \frac{\sqrt{3}}{2} y_{\nu} \frac{v_4}{\sqrt{2}} \\ 0 & M_{\Sigma} \end{pmatrix}}   {\begin{pmatrix} l_R \\ (\Sigma_L^+)^C \end{pmatrix}} + h.c.
\eeq
%%%%%%%%%%%%%%%%%%%%%%%%%%%%%%
Diagonalization of the mass matrices leads to the following mixing matrices \cite{Grimus:2000vj,Kumericki:2012bh}
%%%%%%%%%%%%%%%%%%%%%%%%%%%%%%
\beq
{\begin{pmatrix} \nu_L^C \\ \Sigma^0_R \end{pmatrix}}={\begin{pmatrix} U_{PMNS}^* & \frac{1}{\sqrt{2}} V^* \\ -\frac{1}{\sqrt{2}} V^T U_{PMNS}^*  & \mathds{1} \end{pmatrix}}   {\begin{pmatrix} \tilde{\nu}_L^C \\ \tilde{\Sigma}^0_R
\end{pmatrix}},
\label{eq:mix_nu}
\eeq
\beq
{\begin{pmatrix} l_L \\ (\Sigma^+_R)^C \end{pmatrix}}={\begin{pmatrix} \mathds{1} & \frac{\sqrt{3}}{2} V \\ -\frac{\sqrt{3}}{2} V^{\dagger}  & \mathds{1} \end{pmatrix}}   {\begin{pmatrix} \tilde{l}_L \\ (\tilde{\Sigma}^+_R)^C \end{pmatrix}}, ~{\begin{pmatrix} l_R \\ (\Sigma^+_L)^C \end{pmatrix}}={\begin{pmatrix} \mathds{1} & 0 \\ 0  & \mathds{1} \end{pmatrix}}   {\begin{pmatrix} \tilde{l}_R \\ (\tilde{\Sigma}^+_L)^C \end{pmatrix}}.
\label{eq:mix_l}
\eeq
%%%%%%%%%%%%%%%%%%%%%%%%%%%%%%
%{\bf clarify the content. 
Here, $U_{PMNS}$ is the $3 \times 3$ matrix that diagonalizes the light neutrino masses and $V=\frac{v_4}{\sqrt{2}} y_{\nu}{M_\Sigma}^{-1}$.
%\beq
 %.
%\label{v}
%\eeq
The value of $V$ is constrained by neutrino masses and 
is estimated to be $\sim 3.5 \times 10^{-7}$ \cite{Kumericki:2012bh,Yu:2015pwa}.
The physical states obtained after mixing are denoted by a `$\sim$' over the fields. For simplicity, however, we drop the `$\sim$' hereafter. The $U_{PMNS}$ matrix can be expressed as \cite{ParticleDataGroup:2022pth}
%Matrix $U_{PMNS}$ is expressed as (ref) CHEKKKKKKKK,
%%%%%%%%%%%%%%%%%%%%%%%%%%%%%%
\begin{equation}
U_{\rm PMNS} = \left( \begin{array}{ccc}
c_{12} c_{13} & s_{12} c_{13} & s_{13} e^{-i\delta} \\
-s_{12} c_{23} -c_{12} s_{13} s_{23} e^{i\delta} & c_{12} c_{23} -s_{12} s_{13} s_{23} e^{i\delta} & c_{13} s_{23} \\
s_{12} s_{23} -c_{12} s_{13} c_{23} e^{i\delta} & -c_{12} s_{23} -s_{12} s_{13} c_{23} e^{i\delta} & c_{13} c_{23}
\end{array} \right) \mathrm{diag}(e^{-i \phi/2},e^{-i \phi^\prime/2},1),
\end{equation}
%%%%%%%%%%%%%%%%%%%%%%%%%%%%%%
where $c_{ij}=\cos \theta_{ij}$, $s_{ij}=\sin \theta_{ij}$; $\theta_{12}, \theta_{23}$ and $\theta_{13}$ are the light neutrino mixing angles; $\phi$ and $\phi'$ are Majorana phases; $\delta$ is the Dirac CP violating phase. Data obtained from solar neutrino \cite{SNO:2002tuh,Super-Kamiokande:2001ljr} and atmospheric neutrino searches \cite{Super-Kamiokande:1998kpq} constrain the neutrino masses and the mixing involved. 
As summarized 
in  \cite{ParticleDataGroup:2022pth}, the allowed values of mixing angles and phases are  $\theta_{ij} \in [0,\pi/2]$, $\delta \in [0,\pi]$. 
Considering the large uncertainty in phase values and negligible impact of $\delta$ on determination of Yukawa couplings in seesaw mechanisms \cite{Casas:2001sr}, all phases are assumed to be zero for simplicity. Further, we consider normal hierarchy and use the following best-fit values for $\theta_{ij}$\cite{Esteban:2018azc}
%%%%%%%%%%%%%%%%%%%%%%%%%%%%%%
\bea
\theta_{12}=33.82^\circ~,~\theta_{13}=8.60^\circ,~
\theta_{23}=48.6^\circ.
\eea
%%%%%%%%%%%%%%%%%%%%%%%%%%%%%%
Substituting these values, we obtain
%%%%%%%%%%%%%%%%%%%%%%%%%%%%%%
\begin{equation*}
U_{\rm PMNS} = \left( \begin{array}{ccc}
0.82 & 0.55 & 0.15 \\
-0.46 & 0.48 & 0.74 \\
0.34 & -0.68 & 0.65
\end{array} \right).
\label{upmns}
\end{equation*}
%%%%%%%%%%%%%%%%%%%%%%%%%%%%%%
The gauge couplings of $\Sigma_R$ can be written as
%%%%%%%%%%%%%%%%%%%%%%%%%%%%%%
\bea
\mathcal{L}_{gauge,\Sigma} &\supset& (e A_\mu + g c_w Z_\mu) ~\bigg[ 2 (\bar{\Sigma}^{++}_R \gamma^\mu \Sigma^{++}_R + \bar{\Sigma}^{++}_L \gamma^\mu \Sigma^{++}_L) + \bar{\Sigma}^{+}_R \gamma^\mu \Sigma^{+}_R + \bar{\Sigma}^{+}_L \gamma^\mu \Sigma^{+}_L \bigg] - \nonumber \\
&& g ~\bigg\{ \bigg[ \sqrt{2} (\bar{\Sigma}^{+}_R \gamma^\mu \Sigma^{++}_R - \bar{\Sigma}^{+}_L \gamma^\mu \Sigma^{++}_L) + \sqrt{3} (\bar{\Sigma}^{0}_R \gamma^\mu \Sigma^{+}_R - \bar{(\Sigma^{0}_R)^C} \gamma^\mu \Sigma^{+}_L) \bigg] W^-_\mu + h.c. \bigg\}
\label{eq:gauge_sigma}
\eea
%%%%%%%%%%%%%%%%%%%%%%%%%%%%%%
We combine the similarly charged components to form Dirac fermions
%%%%%%%%%%%%%%%%%%%%%%%%%%%%%%
\beq
\Sigma^{++} = \Sigma_{1,R}^{++} + (\Sigma_{2,R}^{--})^C \equiv \Sigma^{++}_R + \Sigma^{++}_L, ~~~~
\Sigma^{+} = \Sigma_{1,R}^{+} + (\Sigma_{2,R}^{-})^C \equiv \Sigma^{+}_R + \Sigma^{+}_L.
\eeq
%%%%%%%%%%%%%%%%%%%%%%%%%%%%%%
$\Sigma^{0}\equiv\Sigma^{0}_R$ remains as the neutral Majorana fermion. Masses of each fermion component are given by $M_{\Sigma}$.
The corresponding SM interactions are
%%%%%%%%%%%%%%%%%%%%%%%%%%%%%%
\bea
\mathcal{L}_{gauge,SM} &\supset& \frac{g}{c_w} \bigg[ \bar{\nu}_L \gamma^\mu (\frac{1}{2}) \nu_L + \bar{l}_L \gamma^\mu (s_w^2-\frac{1}{2}) l_L \bigg] Z_\mu + \frac{g}{\sqrt{2}} \bigg[ \bar{l}_L \gamma^\mu \nu_L W^-_\mu + h.c. \bigg].
\label{eq:gauge_SM}
\eea
%%%%%%%%%%%%%%%%%%%%%%%%%%%%%%
Substituting the mass eigenstates from Eqs.~\ref{eq:mix_nu} and 
\ref{eq:mix_l} in Eqs.~\ref{eq:gauge_sigma} and \ref{eq:gauge_SM} leads to the following interaction Lagrangian, which describes the interaction among the quintuplet fermions, SM leptons, and gauge bosons.
%%%%%%%%%%%%%%%%%%%%%%%%%%%%%%
\bea
\mathcal{L}_{NC} &=& \frac{g}{cw} ~\Bigg\{ ~\bar{\nu} \gamma^{\mu} \Big[ \frac{1}{2\sqrt{2}}  U^{\dagger}V P_L - \frac{1}{2\sqrt{2}} U^T V^* P_R \Big] \Sigma^0 Z_\mu \nonumber \\
&& + \Big[ ~\bar{l^C} \gamma^{\mu} (\frac{\sqrt{3}}{4} V^* P_R) \Sigma^+ Z_{\mu} + h.c. \Big] \Bigg\},
\eea
%%%%%%%%%%%%%%%%%%%%%%%%%%%%%%
and
%%%%%%%%%%%%%%%%%%%%%%%%%%%%%%
\bea
\mathcal{L}_{CC} &=& g ~\Bigg\{ ~\bar{\nu} \gamma^{\mu} \bigg[ \frac{\sqrt{3}}{2\sqrt{2}} U^T V^* P_R + \frac{\sqrt{3}}{\sqrt{2}} U^{\dagger} V P_L \bigg] \Sigma^+ \nonumber \\
&& - ~\bar{l} \gamma^{\mu} (V P_L) \Sigma^0 + \bar{l^C} \gamma^{\mu} \bigg( \sqrt{\frac{3}{2}} V^* P_R \bigg) \Sigma^{++} \Bigg\} ~W^-_{\mu} + h.c.,
\eea
%%%%%%%%%%%%%%%%%%%%%%%%%%%%%%
where, $U$ stands for the $U_{PMNS}$. 
As discussed before, the value of $V$ is constant and therefore, 
the decay of quintuplets to gauge bosons is independent of the Yukawa coupling.
 
%%%%%%%%%%%%%%%%%%%%%%%%%%%%%%%%%%%
\section{Production and Decay}
%%%%%%%%%%%%%%%%%%%%%%%%%%%%%%%%%%%%%%%%%%%%%%%%%%%%%

In this section, we study the production and decay modes of the quartet scalars at $14$ TeV LHC and discuss the possible collider signatures in the {\em fermiophobic} and {\em fermiophilic} scenarios. We also discuss the existing limits on the quartet scalars from various collider studies.

%%%%%%%%%%%%%%%%%%%%%%%%%%%%%%%%%%%%%%%%%
\subsection{Production at the LHC}
\begin{figure}[h!]
\begin{center}
\includegraphics[width=7cm,height=5.8cm]{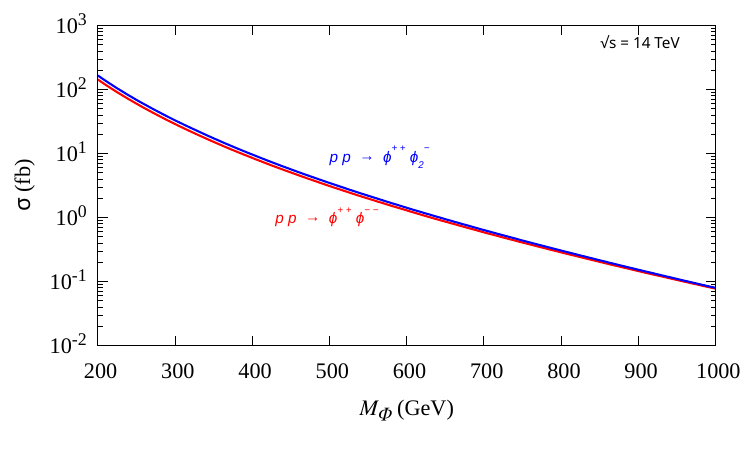}
\includegraphics[width=7cm,height=5.8cm]{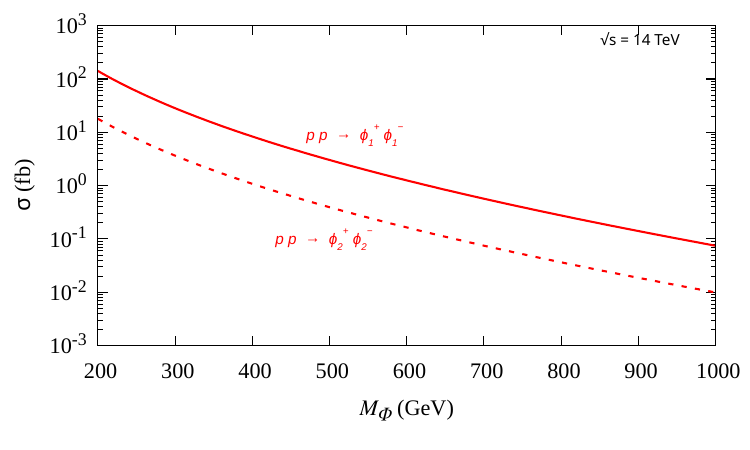}
\caption{(left) Cross sections for the doubly charged scalar pair production and associated production, (right) pair production cross section of the singly charged scalar as a function of their mass at 14 TeV LHC.}
\label{fig:prod_LHC}
\end{center}
\end{figure}

%%%%%%%%%%%%%%%%%%%%%%%%%%%
The charged scalars are produced in $pp$ collisions via SM gauge bosons in the $s$-channel. %, as shown in Fig.~\ref{fig:feyn_FB}. 
We take particular interest in the pair and associated productions of the charged scalars due to their large cross sections compared to the production cross sections (pair or associated) of the neutral scalars viz., 
\bea
&&pp \rightarrow \phi^{++}\phi^{--}, ~~~~~pp \rightarrow \phi^{++}\phi_2^- + h.c.; \nonumber\\
&&pp \rightarrow \phi_1^+\phi_1^-, ~~~~~~~~pp \rightarrow \phi_2^+\phi_2^-.
\eea
We plot these cross sections ($\sigma$) 
as a function of the scalar mass at 14 TeV LHC in Fig.~\ref{fig:prod_LHC}.
The production cross section for $\phi_1^{+}\phi_1^{-}$ differs from the production cross section for $\phi_2^{+}\phi_2^{-}$, and is larger by a factor of $\sim 10$.
%=======================
%=======================
For the doubly charged scalar, the associated production ($\phi^{++}\phi_2^- + h.c.$) is larger than the pair production. There is no associated production of doubly charged scalar with $\phi_1$ because the gauge interaction in Eq.~\ref{eq:prod_gauge} does not couple $\phi^{\pm\pm}$ (isospin $3/2$) with $\phi_1^\pm$ (isospin $-3/2$). Further, the associated production takes place via a $W$-boson as a mediator. Since $\sigma(pp \rightarrow W^+) > \sigma(pp \rightarrow W^-)$, the production cross section of $\phi_2^+\phi^{--}$ is less than that of $\phi^{++}\phi_2^-$.

%%%%%%%%%%%%%%%%%%%%%%%%%%%%%%%%%%%%%%%%%%%%%%
\subsection{Fermiophobic and Fermiophilic Decays}
%%%%%%%%%%%%%%%%%%%%%%%%%%%%%%%%%%%%%%%%%%%%%%%%%%
When $M_{\Phi}<M_\Sigma$, the quadruplet scalars decay only to the gauge bosons and behave as purely {\it fermiophobic} \cite{Nomura:2017abu,Kumar:2019tat,Kumar:2021umc}. The decay modes are
%%%%%%%%%%%%%%%%%%%%%%%%%%%%%%
\bea
\phi^{\pm}_{1,2} &\rightarrow & W^{\pm} Z, \n \\
\phi^{\pm \pm} &\rightarrow & W^{\pm} W^{\pm}, \n \\
\phi^{0} &\rightarrow & W^{+} W^{-}.
\label{eq:phi_decays_1}
\eea
%%%%%%%%%%%%%%%%%%%%%%%%%%%%%%
On the other hand, when the scalars are heavier than the quintuplet fermions \emph{i.e.,} $M_{\Phi}>M_\Sigma$, then the scalars have an additional decay mode whereby they decay to quintuplet fermions via Eq.~\ref{eq:yukawa}.
These {\it fermiophilic} decay modes are
%%%%%%%%%%%%%%%%%%%%%%%%%%%%%%
\bea
\phi_1^{\pm} &\rightarrow& \Sigma^{0} l^{\pm} /   \Sigma^{\pm} \nu, \n \\
\phi_2^{\pm} &\rightarrow&  \Sigma^{\pm} \nu / \Sigma^{\pm \pm} l^{\mp}, \n \\
\phi^{\pm \pm}  &\rightarrow & \Sigma^{\pm \pm} \nu / \Sigma^{\pm} l^{\pm}, \n \\
\phi^{0} &\rightarrow & \bar{\nu}_l \Sigma^{0} / \bar{l} \Sigma^{-}.
\label{eq:phi_decays_2}
\eea
%%%%%%%%%%%%%%%%%%%%%%%%%%%%%%
\begin{figure}[h!]
\begin{center}
\begin{subfigure}[b]{0.45\textwidth}
\includegraphics[width=7.5cm,height=6cm]{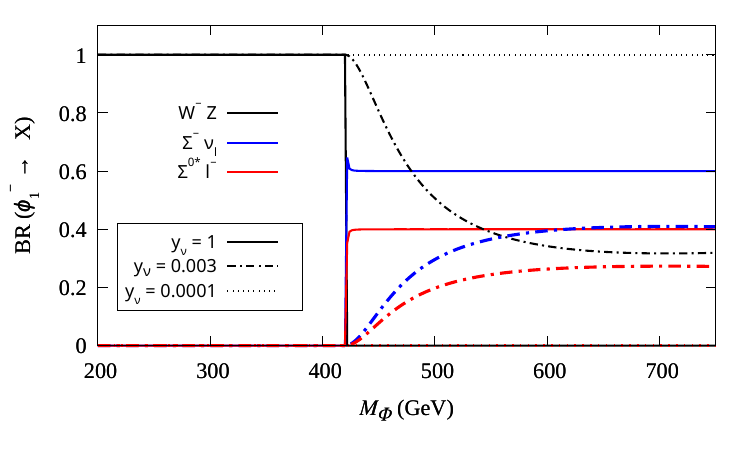}
         \caption{BR ($\phi_1^{-}$)}
         \label{fig:br_phica}
     \end{subfigure}
\begin{subfigure}[b]{0.45\textwidth}
\includegraphics[width=7.5cm,height=6cm]{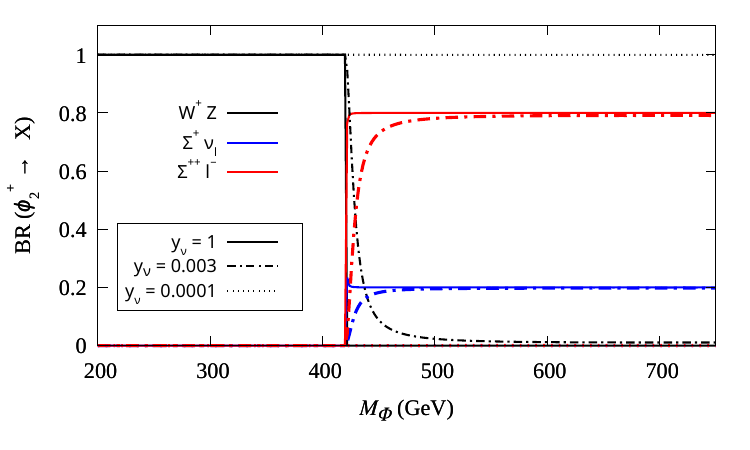}
         \caption{BR ($\phi_2^+$)}
         \label{fig:br_phicb}
     \end{subfigure}
\begin{subfigure}[b]{0.45\textwidth}
\includegraphics[width=7.5cm,height=6cm]{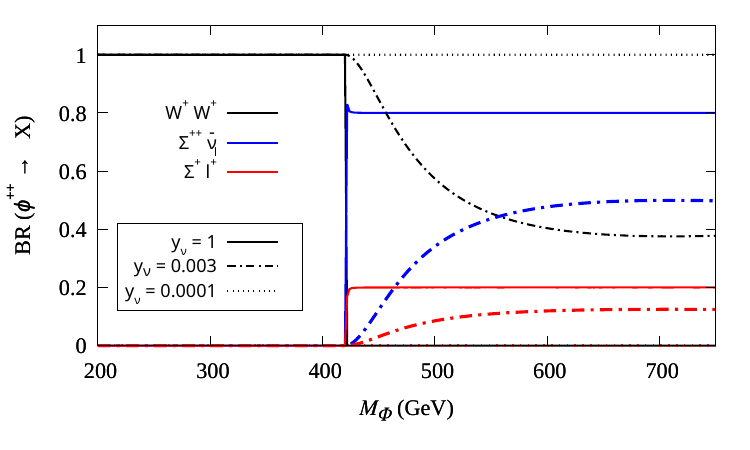}
         \caption{BR ($\phi^{++}$)}
         \label{fig:br_phicc}
     \end{subfigure}
\caption{Branching ratios for the singly  ($\phi^{\pm}$) and the doubly charged ($\phi^{\pm\pm}$) scalars considering three different values of the Yukawa coupling, as indicated in graphs. 
Here, $l=e,\mu,\tau$. The quartet and quintuplet masses are fixed at $M_\Phi=500$ GeV and $M_\Sigma=420$ GeV. 
}
\label{fig:br_phi}
\end{center}
\end{figure}
%%%%%%%%%%%%%%%%%%%%%%%%%%%%%%
The branching ratios (BR) of $\phi$s are depicted in Fig.~\ref{fig:br_phicc} as a function of $M_{\Phi}$ while $M_{\Sigma}$ is fixed at $420$ GeV. When $M_{\Phi}<M_\Sigma$, $\Phi$ decays to gauge bosons only.
The {\em fermiophilic} modes open up when $M_{\Phi} > M_{\Sigma}$.
In the {\em fermiophilic} region, we vary the Yukawa coupling and choose $y_{\nu}^{e} = y_{\nu}^{\mu} =y_{\nu}^{\tau}$ for simplicity.
The corresponding values of VEV are in the range (0.0002 -1) GeV.
We observe that at large values of the $y_\nu$, the scalars prefer to decay to the quintuplet fermion and SM leptons. Hence, the decays are purely {\it fermiophilic}. If we lower the value of the Yukawa coupling, the branching ratio to the gauge bosons starts to dominate. For the singly charged scalar ($\phi_1^{\pm}$), the {\em fermiophilic} and {\em fermiophobic} branching ratios are comparable when $Y_\nu=0.003$. However, for $\phi_2^{\pm}$, the {\em fermiophilic} decay is still preferred when $Y_\nu=0.003$.  
Hence, we can say that the two singly charged scalars, $\phi_1$ and $\phi_2$ show different decay patterns at intermediate values of $y_\nu$.
At much smaller values of $Y_\nu$, the decays of both the scalars are {\em fermiophobic}.

In the {\em fermiophilic} decay modes, the quintuplet fermions further decay to SM gauge bosons and leptons via the following modes:
%%%%%%%%%%%%%%%%%%%%%%%%%%%%
\bea
\Sigma^{0} &\rightarrow & l^- W^+ ,\nu_l Z, \n \\
\Sigma^{\pm} &\rightarrow & l^{\pm} Z , \nu W^{\pm}, \n \\
\Sigma^{\pm \pm} &\rightarrow & W^{\pm} l^{\pm}.
\label{eq:sigmadecay}
\eea
%%%%%%%%%%%%%%%%%%%%%%%%%%%%
As discussed in Sec.~\ref{sec:model}, the branching ratios of $\Sigma$s are constant with respect to variation in Yukawa coupling and found to be $1$ for $\Sigma^0 \rightarrow W^+l^-$, $\Sigma^{\pm} \rightarrow W^+\nu_l$, and $\Sigma^{\pm\pm} \rightarrow W^+l^+$. Hence, while studying the collider signature in the next section, we consider these decay modes only.
%==================================
\subsection{Existing Collider Searches}
%====================================
 In the model under inspection, the scalars decay either to the gauge bosons directly or to the quintuplet fermions which subsequently decay to gauge bosons, as we have discussed in the previous section. Before discussing possible signatures at the LHC, let us discuss the existing limits from collider experiments on charged scalars. The direct search limit on charged Higgs comes from LEP which is around 80 GeV \cite{ALEPH:2013htx}.
 Both ATLAS and CMS have conducted extensive searches targeting multiple production and decay modes of the charged scalars. A recent summary of these searches is provided in \cite{Tao:2021tzs,ATLAS:2024itc,Horii:2024qiw,Banerjee:2025qkb}.  %CHEKKKKKKKK. 
Most LHC searches target single production of the charged scalars via vector-boson fusion and have put limits in cases where the scalar decays to gauge bosons\cite{ATLAS:2022zuc,ATLAS:2023sua,CMS:2021wlt,ATLAS:2018ceg}. In this case, a lower limit of $\sim 200$ GeV is placed on the masses of charged scalars.
Limits also exist on production cross sections of charged scalars when they directly decay to SM fermions \cite{CMS:2019bfg,ATLAS:2012nhc,ATLAS:2018gfm,ATLAS:2018ntn,ATLAS:2015nkq,ATLAS:2022pbd} or specifically, to quarks \cite{ATLAS:2013uxj}. These searches have placed a lower limit of $\sim 400$ GeV on the charged scalar masses.

Ref.~\cite{ATLAS:2021jol} has studied the doubly charged scalar pair production as well as associated production of singly  and doubly charged scalars, with each scalar decaying to gauge bosons and limits are placed from the searches in 4 leptons, 3 leptons + $\geq$ 2 jets and 2 leptons +$\geq$ 3 jets final states. As no excess has been observed, limits have been placed on the cross section $\times$ BR (See Fig. 9 of \cite{ATLAS:2021jol}).

One key aspect is that ATLAS and CMS searches consider that the scalars decay to the SM gauge bosons or fermions directly. However, in the model under study, the singly  and doubly charged scalars are part of a quadruplet. They decay to SM gauge bosons ({\em fermiophobic} scenario) or to quintuplet fermions ({\em fermiophilic} scenario) which further decay to SM particles. Hence, there are a few scenarios where the experimental limits are applicable, which we discuss next.
%============================================
\subsection{Fermiophobic and Fermiophilic Signatures}
%============================================
Let us consider the {\em fermiophobic} scenario when $M_{\Phi} > M_{\Sigma}$ and the coupling of the scalars with the quintuplet fermion is small, $y_\nu<0.005$. In this situation, the branching ratio of the scalars to the SM fermions are negligible. This situation gives rise to the final states with gauge bosons
%%%%%%%%%%%%%%%%%%%%%%%%%%%%
\bea
p p ~\rightarrow~ \phi_1^{+} \phi_1^{-} &\rightarrow& W^+ Z ~W^- Z, \\
p p \rightarrow \phi^{++} \phi^{--} &\rightarrow& W^{+} W^{+} W^{-} W^{-}, \\
p p \rightarrow \phi^{++} \phi_2^{-} &\rightarrow& 
     W^{+} W^{+} ~W^{+} Z.  
%\label{eq:sig_fermiophobic_3}
\eea
%%%%%%%%%%%%%%%%%%%%%%%%%%%%
We present the possible final states and the values of cross section $\times$ BR in Table~\ref{tab:sigmabr_1} for different masses of the scalars. 
Some of the processes also get limits from ATLAS study~\cite{ATLAS:2021jol}. 
We have shown these limits in parentheses, in Table~\ref{tab:sigmabr_1}, comparing processes with the same initial and final states, in our model and the ATLAS study. 
We cite the limits from Ref. \cite{ATLAS:2021jol} in $4l$, $3l+ \geq 2j$ and $2l+ \geq 3j$ channels. Note that the limit in  $2l+ \geq 3j$ channel is applicable only when the two leptons have the same sign. We also did not include any channel with jet multiplicity greater than 4 because these channels suffer from large SM background.
%%%%%%%%%%%%%%%%%%%%%%%%%%%%%%%%%%%%%%%%%
\begin{table}
    \centering
    \begin{tabular}{|cc|c|c|c|}
         \hline
         \multicolumn{2}{|c|}{\multirow{ 2}{*}{Process}} & \multicolumn{3}{c|}{Cross section$\times$ BR (fb)}\\ 
         %\hline
         ~&~ & {$300 ~GeV$} & {$500 ~GeV$} & {$800 ~GeV$} \\
         \hline
%1& $2l$+ MET &&&\\
%%%%%%%%%%%%%%%%%%%%%%%%%%%%%%%%%%%%%%%%%%%%%%%%%%
 $p p \rightarrow \phi_1^{+} \phi_1^{-}$: 
 & $6l+ \cancel{E}_T$ & $0.01$ & $0.001$ & $0.0001$\\
~&$5l+2j+ \cancel{E}_T$ & $0.054$ & $0.0055$ & $0.0005$\\
~&$4l+4j+ \cancel{E}_T$ & $0.07$ & $0.007$ & $0.0006$\\
~&$4l+2j+ \cancel{E}_T$ & $0.184$ & $0.019$ & $0.0016$\\
\hline
%%%%%%%%%%%%%%%%%%%%%%%%%%%%%%%%%%%%%%%%%%%%%%%%%%
$ p p \rightarrow \phi^{++} \phi^{--}$: 
& $4l + \cancel{E}_T$ & $0.114 ~(\bf0.044)$ & $0.0117 ~(\bf0.026)$ & $0.001$\\
~&$3l+2j+\cancel{E}_T$ & $1.17 ~(\bf 0.45)$ & $0.12 ~(\bf0.26)$ & $0.01$\\ 
~&$2l_{SC}+4j+\cancel{E}_T$ & $1.5~(\bf 0.58)$ & $0.15~(\bf 0.34)$ & $0.013$\\ 
\hline
%%%%%%%%%%%%%%%%%%%%%%%%%%%%%%%%%%%%%%%%%%%%%%%%%%
$p p \rightarrow \phi^{++} \phi_2^{-}$: 
& $5l$ & $0.0538$ & $0.0059$ & $0.0006$\\
~& $4l+2j+\cancel{E}_T$ & $0.4231$ & $0.046$ & $0.0044$\\
~& $3l+4j+\cancel{E}_T$ & $1.07~(\bf 0.75)$ & $0.117~(\bf 0.42)$ & $0.011$\\
~& $3l+2j+\cancel{E}_T$ & $0.473~(\bf 0.33)$ & $0.0518 ~(\bf 0.18)$ & $0.0049$\\
~& $2l_{SC}+4j+\cancel{E}_T$ & $1.2115~(\bf 0.85)$ & $0.133~(\bf 0.47)$ & $0.0125$\\
%%%%%%%%%%%%%%%%%%%%%%%%%%%%%%%%%%%%%%%%%%%%%%%%%%

\hline
 \end{tabular}
   \caption{{\em fermiophobic} scenario:
   Cross section $\times$ BR for 
   various final states at the $14$ TeV LHC, for $M_\Phi=300,~500$ and $800$ GeV, with $M_\Sigma=M_\Phi-80$ GeV. 
   We use $y_\nu=0.0001$. 
   The values in parentheses indicate the existing upper limits from ATLAS~\cite{ATLAS:2021jol}
   .}
    \label{tab:sigmabr_1}
\end{table}
%%%%%%%%%%%%%%%%%%%%%%%%%%%%%%%%%%%%%%%%%%%%%%

We find that the pair production cross section of $\phi_1^{+} \phi_1^{-}$ does not receive any limit from the LHC. The doubly charged scalar with mass 300 GeV is excluded from the ATLAS search in all three channels when produced in pair. However, doubly charged scalars with mass 500 GeV are not excluded in these channels.
In the associated production mode also, the 300 GeV mass is excluded in $3l+4j$, $3l +2j$ and $2l_{SC}+4j$ channels from $3l ~+~ \geq 2j$ and $2l ~+~ \geq 3j$ channel searches of ATLAS. The multilepton channels do not face any constraints from the LHC, but the cross sections in these channels (with four or more leptons) remain significantly small.

There exists no limit from the LHC on the remaining channels, involving four or more leptons. Note that these final state leptons originate from $W/Z$ boson decays and thus, these channels experience suppression due to small branching ratio. Only the channel with four leptons and two jets offers a better cross section $\times$ BR.  However, as two of the leptons come from the decay of $W$ boson, the 
reconstruction of the quartet scalar or the quintuplet fermion is not straightforward. In the {\em fermiophilic} scenario, which we discuss next, the leptons are produced directly from the decay of the scalar or fermion, resulting in a large cross section $\times$ BR. This also facilitates the reconstruction of the BSM particles.

%%%%%%%%%%%%%%%%%%%%%%%%%%%%%%%%%%%%%%%%
%\subsection{Fermiophilic Signatures}
%%%%%%%%%%%%%%%%%%%%%%%%%%%%%%%%%%%%%%%%%
We consider the {\em fermiophilic} scenario with $M_{\Phi} > M_{\Sigma}$ and $y_\nu=1$. These {\em fermiophilic} signals are very different from the existing searches at the LHC.
The existence of an intermediate state in the {\em fermiophilic} scenario results in a high number of particles in the final state, producing multi-lepton final states with or without multiple jets.
If we consider the pair production of $\phi_1^{+} \phi_1^{-} $ and dominating decay modes of $\phi_1^{\pm}$ and $\Sigma^0$, then final states are rich in leptons and $W$ bosons:
%%%%%%%%%%%%%%%%%%%%%%%%%%%%%%%%%%%%%
\bea 
\label{eq:sig_fl_1}
p p \rightarrow \phi_1^{+} \phi_1^{-} &\rightarrow&
    \Sigma^{0} l^{+} ~\Sigma^{0} l^{-} ~~\rightarrow W^+ l^- l^{+} W^- l^+ l^{-}.
\eea
%%%%%%%%%%%%%%%%%%%%%%%%%%%%%%%%%%%%%
The other possibilities are $p p \rightarrow \phi_1^{+} \phi_1^{-} \rightarrow  \Sigma^{+} \nu ~\Sigma^{-}\bar{\nu} $ and $p p \rightarrow \phi_1^{+} \phi_1^{-} \rightarrow  \Sigma^{0} l^+ ~\Sigma^{-}\bar{\nu} $.  We have observed in the previous section that the branching ratio for $\phi_1^+\rightarrow\Sigma^{+} \nu$ is larger than $\phi_1^+\rightarrow\Sigma^{0} l^+$. The branching ratio for $\Sigma^{\pm}$ decays to $W^\pm \nu (\bar{\nu})$ is almost 95\%. Hence, the final states from these decay modes will be $W^+W^-$+ MET or $W^\pm W^- l^\mp l^+$ + MET. There will be a large MET in the final states if the $W$s decay leptonically, and thereby, reconstruction of the scalar mass will not be straightforward. On the other hand, $W$s decaying to the jets will lead to final states, which suffer from large SM backgrounds.

The other possible {\em fermiophilic} channels are
%%%%%%%%%%%%%%%%%%%%%%%%%%%%%%%%%%%%%
\bea   
\label{eq:sig_fl_2}
   p p \rightarrow \phi^{++} \phi^{--} &\rightarrow& 
    \Sigma^{++} \nu ~\Sigma^{--}\bar{\nu} ~~\rightarrow W^+ l^+ \nu W^- l^- \bar{\nu}, \\
\label{eq:sig_fl_2_b}
     p p \rightarrow \phi^{++} \phi_2^{-} &\rightarrow&
     \Sigma^{++} \nu ~\Sigma^{--}l^{+} \rightarrow W^+ l^+ \nu W^- l^- l^+. 
\eea
%%%%%%%%%%%%%%%%%%%%%%%%%%%%%%%%%%%%%
The cross section $\times$ BR in different channels are presented in Table~\ref{tab:sigmabr_2}. 
We do not consider the pair production of $\phi_2^+\phi_2^-$ because the cross section is one order of magnitude smaller.
Some of the processes also get a limit from the ATLAS study \cite{ATLAS:2021jol}, which are shown in parentheses. However, these limits are only meant for a preliminary comparison. Even though the final states are the same, the exact signal topology, particularly in the  {\em fermiophilic} scenario, is different from that in the ATLAS study.

We observe from~\cite{ATLAS:2021jol} that the 1st process (Eq.~\ref{eq:sig_fl_1}) does not get any limits from the LHC but the cross section $\times$ BR is low compared to the other channels.
The doubly charged scalars with 500 GeV are excluded by the ATLAS search in $3l+\geq 2j$ and $4l$ channels
\footnote{ ATLAS limit does not apply to $2l+4j$ state here because ATLAS considers only same charge leptons, whereas in our study we have two oppositely charged leptons.}. The $2l+4j$ channel with two opposite sign leptons suffers from a very high SM background.
Also, in the associated production mode, scalars with 300 GeV mass are excluded in the $3l+4j$ channel.
%%%%%%%%%%%%%%%%%%%%%%%%%%%%%%%%%%%%%%%%%
\begin{table}[htb!]
    \centering
    \begin{tabular}{|cc|c|c|c|}
         \hline
         \multicolumn{2}{|c|}{\multirow{ 2}{*}{Process}} & \multicolumn{3}{c|}{Cross section$\times$ BR (fb)}\\ 
         %\hline
         ~ & ~ & {$300 ~GeV$} & {$500 ~GeV$} & {$800 ~GeV$} \\
         \hline
 $p p \rightarrow \phi_1^{+} \phi_1^{-}$: & $6l+ \cancel{E}_T$ & $0.083$ & $0.0085$ & $0.00074$\\
~&$5l$+2j & $0.4264$ & $0.044$ & $0.0038$\\
~&$4l$+4j& $0.5467$ & $0.056$ & $0.00489$\\
\hline
$ p p \rightarrow \phi^{++} \phi^{--}$: & $4l+\cancel{E}_T$ & $27.26 (\bf 0.044)$ & $2.8 (\bf 0.026)$ & $0.24$\\
~&$3l+2j+\cancel{E}_T$& $3.577 (\bf 0.45)$ & $0.3675 (\bf 0.26)$ & $0.0315$\\
~&$2l+4j+\cancel{E}_T$& $4.542$ & $0.467$ & $0.04$\\
\hline
$p p \rightarrow \phi^{++} \phi_2^{-}$: & $5l$ & $0.838$ & $0.0918$ & $0.0086$\\
~& $4l+2j+\cancel{E}_T$ & $4.3$ & $0.471$ & $0.0443$ \\
~& $3l+4j+\cancel{E}_T$ & $5.514 (\bf 0.75)$ & $0.6035 (\bf 0.418)$ & $0.0568$ \\

\hline
 \end{tabular}
   \caption[x]{
   {\em fermiophilic} scenario: 
   Cross section $\times$ BR for 
   various final states at the $14$ TeV LHC, for $M_\Phi=300,~500$ and $800$ GeV, with $M_\Sigma=M_\Phi-80$ GeV. 
   We use parameter values $y_\nu=1$ and $v_4=0.0002$. The values in parentheses indicate the existing upper limits from ATLAS~\cite{ATLAS:2021jol}.}
    \label{tab:sigmabr_2}
\end{table}
%%%%%%%%%%%%%%%%%%%%%%%%%%%%%%%%%%%%%%%%%%%%%%
The multilepton final states with 4 to 5 leptons and 2 or more jets are not excluded by the LHC searches, but they suffer from a very small branching ratio.
Note that, in the {\em fermiophilic} signals, all or most of the SM leptons originate from the decay of the scalars or the quintuplet fermions, while the jets come from the decays of $W/Z$ bosons. Hence, the branching ratio is larger compared to the {\em fermiophobic} scenario (Table~\ref{tab:sigmabr_2}). The mass of the charged scalars and the quintuplet fermions can be reconstructed easily in the {\em fermiophilic} scenario. 

Considering all the current bounds on the masses/couplings of the model under consideration, in the next section, we study the pair production  $\phi_1^{+} \phi_1^{-}$ and associated production  $\phi^{++} \phi_2^{-}$ with final state signature involving multiple leptons (at least 4) and at least 2 jets at the LHC.
%------------------------------
%%%%%%%%%%%%%%%new%%%%%%%%%%%%
%\newpage
\section{Analysis and Results}
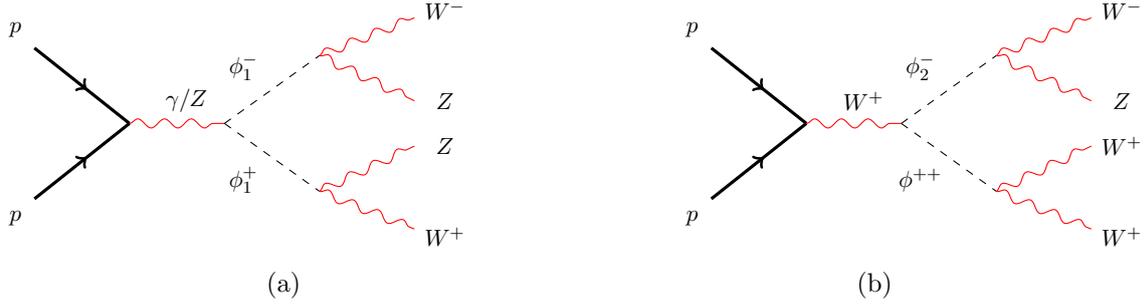
\begin{figure}[h!]
  \centering
\begin{subfigure}[b]{0.45\textwidth}
%\hspace{1cm}
\begin{tikzpicture}[line width=1.4 pt, scale=1,every node/.style={scale=0.8}]

\draw[fermion,black,very thick] (-6,1) --(-4.75,0);
\draw[fermion,black,very thick] (-6,-1) --(-4.75,0);
%\filldraw[black] (-3.5,0) circle (2pt);

\draw[vector,red,thin] (-4.75,0) --(-3.5,0);

\draw[scalar,black,thin] (-3.5,0) --(-2.25,0.9);
\draw[vector,red,thin] (-2.25,0.9) --(-1,1.4);
\draw[vector,red,thin] (-2.25,0.9) --(-1,0.3);

\draw[scalar,black,thin] (-3.5,0) --(-2.25,-0.9);
\draw[vector,red,thin] (-2.25,-0.9) --(-1,-1.4);
\draw[vector,red,thin] (-2.25,-0.9) --(-1,-0.3);

\node at (-6.25,1.25) {$p$};
\node at (-6.25,-1.25) {$p$};

\node at (-4,0.3) {$\gamma/Z$};

\node at (-3.25,0.75) {$\phi_1^{-}$};
\node at (-3.25,-0.75) {$\phi_1^{+}$};

\node at (-0.6,1.5) {$W^-$};
\node[scale=1] at (-0.6,0.3) {$Z$};
\node[scale=1] at (-0.6,-0.3) {$Z$};
\node at (-0.6,-1.5) {$W^+$};

\end{tikzpicture}
         \caption{$~$}
         \label{fig:feyn_FB_1}
     \end{subfigure}
%\begin{subfigure}[b]{0.45\textwidth}
%\hspace{1cm}
%\input{production/feyn_FB_2.tex}
%         \caption{$~$}
%         \label{fig:feyn_FB_2}
%     \end{subfigure}
\begin{subfigure}[b]{0.45\textwidth}
\hspace{1cm}
\begin{tikzpicture}[line width=1.4 pt, scale=1,every node/.style={scale=0.8}]

\draw[fermion,black,very thick] (-6,1) --(-4.75,0);
\draw[fermion,black,very thick] (-6,-1) --(-4.75,0);
%\filldraw[black] (-3.5,0) circle (2pt);

\draw[vector,red,thin] (-4.75,0) --(-3.5,0);

\draw[scalar,black,thin] (-3.5,0) --(-2.25,0.9);
\draw[vector,red,thin] (-2.25,0.9) --(-1,1.4);
\draw[vector,red,thin] (-2.25,0.9) --(-1,0.3);

\draw[scalar,black,thin] (-3.5,0) --(-2.25,-0.9);
\draw[vector,red,thin] (-2.25,-0.9) --(-1,-1.4);
\draw[vector,red,thin] (-2.25,-0.9) --(-1,-0.3);

\node at (-6.25,1.25) {$p$};
\node at (-6.25,-1.25) {$p$};

\node at (-4,0.3) {$W^+$};

\node at (-3.25,0.75) {$\phi_2^{-}$};
\node at (-3.25,-0.75) {$\phi^{++}$};

\node at (-0.6,1.5) {$W^-$};
\node[scale=1] at (-0.6,0.3) {$Z$};
\node[scale=1] at (-0.6,-0.3) {$W^+$};
\node at (-0.6,-1.5) {$W^+$};

\end{tikzpicture}
         \caption{$~$}
         \label{fig:feyn_FB_3}
     \end{subfigure}
  \caption{Feynman diagrams for the pair production and associated production of the charged scalars at $pp$ collision and their decays in the  {\em fermiophobic} scenario.}
\label{fig:feyn_FB}
\end{figure}

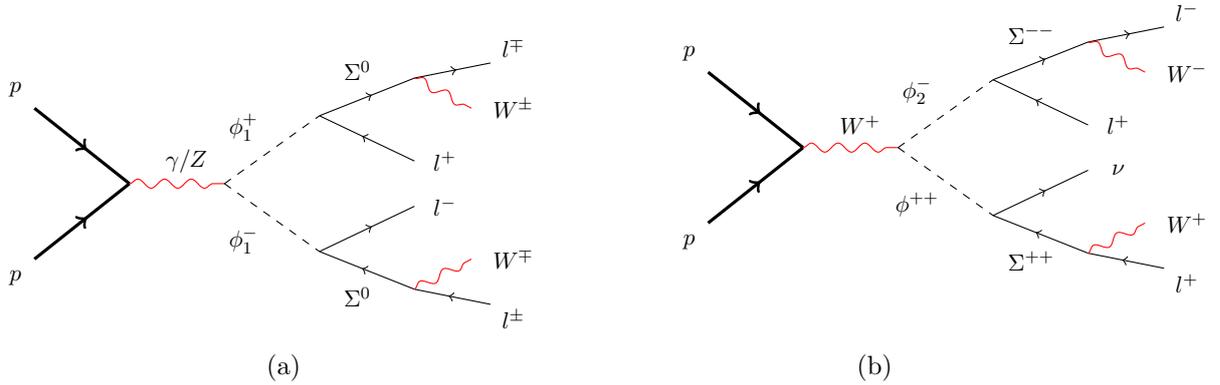
\begin{figure}[h!]
  \centering
\begin{subfigure}[b]{0.45\textwidth}
%\hspace{1.2cm}
\begin{tikzpicture}[line width=1.4 pt, scale=1,every node/.style={scale=0.8}]

\draw[fermion,black,very thick] (-6,1) --(-4.75,0);
\draw[fermion,black,very thick] (-6,-1) --(-4.75,0);
%\filldraw[black] (-3.5,0) circle (2pt);

\draw[vector,red,thin] (-4.75,0) --(-3.5,0);

\draw[scalar,black,thin] (-3.5,0) --(-2.25,0.9);
\draw[fermion,black,thin] (-2.25,0.9) --(-1,1.4);
\draw[fermion,black,thin] (-1,0.3) --(-2.25,0.9);
\draw[fermion,black,thin] (-1,1.4) --(0,1.6);
\draw[vector,red,thin] (-1,1.4) --(-0.25,1);
%\draw[fermion,black,thin] (-0.25,1) --(0.25,0.75);
%\draw[fermion,black,thin] (-0.25,1) --(0.25,1.25);

\draw[scalar,black,thin] (-3.5,0) --(-2.25,-0.9);
\draw[fermion,black,thin] (-2.25,-0.9) --(-1,-0.3);
\draw[fermion,black,thin] (-1,-1.4) --(-2.25,-0.9);
\draw[fermion,black,thin] (0,-1.6) --(-1,-1.4);
\draw[vector,red,thin] (-1,-1.4) --(-0.25,-1);
%\draw[fermion,black,thin] (-0.25,-1) --(0.25,-0.75);
%\draw[fermion,black,thin] (-0.25,-1) --(0.25,-1.25);

\node at (-6.25,1.25) {$p$};
\node at (-6.25,-1.25) {$p$};

\node at (-4,0.3) {$\gamma/Z$};

\node at (-3.25,0.75) {$\phi_1^{+}$};
\node at (-3.25,-0.75) {$\phi_1^{-}$};

\node at (-1.75,1.5) {$\Sigma^{0}$};
\node[scale=1] at (-0.6,0.3) {$l^+$};
\node[scale=1] at (-0.6,-0.3) {$l^-$};
\node at (-1.75,-1.5) {$\Sigma^{0}$};

\node at (0.3,1.8) {$l^\mp$};
\node at (0.3,1) {$W^\pm$};
\node at (0.3,-1) {$W^\mp$};
\node at (0.3,-1.8) {$l^\pm$};

\end{tikzpicture}
        
			            
         \caption{$~$}
         \label{fig:feyn_FL_1}
     \end{subfigure}
%\begin{subfigure}[b]{0.45\textwidth}
%\hspace{1.2cm}
%\input{production/feyn_FL_2.tex}
%         \caption{$~$}
%         \label{fig:feyn_FL_2}
%     \end{subfigure}
\begin{subfigure}[b]{0.45\textwidth}
\hspace{1cm}
\begin{tikzpicture}[line width=1.4 pt, scale=1,every node/.style={scale=0.8}]

\draw[fermion,black,very thick] (-6,1) --(-4.75,0);
\draw[fermion,black,very thick] (-6,-1) --(-4.75,0);
%\filldraw[black] (-3.5,0) circle (2pt);

\draw[vector,red,thin] (-4.75,0) --(-3.5,0);

\draw[scalar,black,thin] (-3.5,0) --(-2.25,0.9);
\draw[fermion,black,thin] (-2.25,0.9) --(-1,1.4);
\draw[fermion,black,thin] (-1,0.3) --(-2.25,0.9);
\draw[fermion,black,thin] (-1,1.4) --(0,1.6);
\draw[vector,red,thin] (-1,1.4) --(-0.25,1);

			\draw[scalar,black,thin] (-3.5,0) --(-2.25,-0.9);
			\draw[fermion,black,thin] (-2.25,-0.9) --(-1,-0.3);
			\draw[fermion,black,thin] (-1,-1.4) --(-2.25,-0.9);
			\draw[fermion,black,thin] (0,-1.6) --(-1,-1.4);
			\draw[vector,red,thin] (-1,-1.4) --(-0.25,-1);

			\node at (-6.25,1.25) {$p$};
			\node at (-6.25,-1.25) {$p$};

\node at (-4,0.3) {$W^+$};

			\node at (-3.25,0.75) {$\phi_2^{-}$};
			\node at (-3.25,-0.75) {$\phi^{++}$};

			\node at (-1.75,1.5) {$\Sigma^{--}$};
			\node[scale=1] at (-0.6,0.3) {$l^+$};
			\node[scale=1] at (-0.6,-0.3) {$\nu$};
			\node at (-1.75,-1.5) {$\Sigma^{++}$};

			\node at (0.3,1.8) {$l^-$};
			\node at (0.3,1) {$W^-$};
			\node at (0.3,-1) {$W^+$};
			\node at (0.3,-1.8) {$l^+$};

			\end{tikzpicture}
         \caption{$~$}
         \label{fig:feyn_FL_3}
     \end{subfigure}
  \caption{Feynman diagrams for the pair production and associated production of the charged scalars at $pp$ collision and their decays in the {\em fermiophilic} scenario. The conjugate processes are also considered.}
\label{fig:feyn_FL}
\end{figure}

%%%%%%%%%%%%%%%%%%%%%%%%%%%%%%%%%%%
The pair production of $\phi_1^{+} \phi_1^{-}$ and associated production  $\phi^{++} \phi_2^{-}$ is shown in Fig.~\ref{fig:feyn_FB} and Fig.~\ref{fig:feyn_FL} for the {\em fermiophobic} and {\em fermiophilic} scenarios, respectively. 
In this section, we discuss the detailed collider analysis corresponding to the {\em fermiophilic} scenario as depicted in Fig.~\ref{fig:feyn_FL}. 
We consider pair production of $\phi_{1}^\pm$ as our first signal process (S1) where $\phi_{1}$ undergoes a cascade decay through the intermediate state $\Sigma^{0}$, resulting in a final state with four charged leptons and two W bosons (Fig.~\ref{fig:feyn_FL_1}). In addition to S1, we also study the associated production of $\phi_{2}^\pm$ with $\phi^{\mp\mp}$, which constitutes our second signal process (S2), resulting in a final state with three charged leptons and two W bosons. Note that, we have considered the charged conjugate process as well.

%----------------------------------------------------
%===============================================
%----------------------------------------------------
%===============================================

%============================================================
\paragraph{Signal Topology}

The processes in Fig.~\ref{fig:feyn_FL} results in a final state containing four/five leptons and two jets when one of the two $W$ bosons decays hadronically while the other decays leptonically.
For S1, both the possibilities are considered. 
There is no other source of MET in S1 apart from the decay of leptonically decaying $W$.  $W$ can be reconstructed from the jets, and can be combined with correct leptons to reconstruct the fermion or scalar. We will discuss the reconstruction later.

However, in S2, if the $W$ originating from the $\phi^{\pm\pm}$ branch decays hadronically, the resulting hadronic system allows the reconstruction of the intermediate $\Sigma^{\pm\pm}$ on that branch. However, the presence of missing transverse energy (MET) from the $\phi^{\pm\pm}$ on the same branch prevents the reconstruction of the parent $\phi^{\pm\pm}$. Furthermore, on the $\phi_{2}^\pm$ side, the presence of MET from the leptonically decaying $W$ again makes it difficult to reconstruct both $\Sigma^{\pm\pm}$ and $\phi_2^\pm$ in that branch.
Conversely, if the $W$ from the $\phi^{\pm\pm}$ branch decays leptonically, all MET is concentrated on the $\phi^{\pm\pm}$ side, rendering both $\Sigma^{\pm\pm}$ and $\phi^{\pm\pm}$ unreconstructible. In this configuration, however, the $\phi_{2}^\pm$ branch contains a hadronically decaying $W$, making it possible to reconstruct both $\Sigma^{\pm\pm}$ and $\phi_2^\pm$. Thus, although the $\phi^{\pm\pm}$ branch cannot be reconstructed in this scenario, the topology still allows the reconstruction of at least one scalar and one fermionic resonance.
For these reasons, in our analysis we focus exclusively on the latter configuration—where the leptonic $W$ originates from the $\phi^{\pm\pm}$ branch—as it provides the most favorable conditions for reconstructing at least one scalar and one heavy fermion state.

\paragraph{Backgrounds:} Since both signal processes produce final states with at least four leptons accompanied by jets and missing transverse energy (MET), the most relevant SM backgrounds are those that can mimic such signatures. Therefore, we consider SM processes containing {\it at least four leptons and two jets}, along with a non-zero source of MET, as potential backgrounds to S1 and S2. The list of dominant SM backgrounds includes $t\bar{t}Z$, $tWZj$, $ZWWjj$, $ZZWjj$, $t\bar{t}WW$, and $t\bar{t}t\bar{t}$, which exhibit similar final-state topologies with cross sections ~\cite{ATLAS:2023dkz, CMS:2020fqz} comparable to (in some cases larger) the signal S1 and S2. A summary of the signal and background processes, along with their relevant decay modes and cross section $\times$ BR, is provided in Table~\ref{tab:xsec}. 

%----------------------------------------
\begin{table}[!htb]
%\hspace{-1.5 cm}
\centering
    \begin{tabular}{|c|c|c|c|c|c|}
    \hline
& Signal &$\sigma\times$ BR (fb) \\ \hline
      $\rm S1:$& $pp \rightarrow \phi^{+}_{1} \phi^{-}_{1};  $ & 0.4264\\
    $M_{\phi}$=300 GeV  & $(\phi_1^{+}\rightarrow \Sigma^{0} l^{+}, \Sigma^{0}\rightarrow W^+ l^-, W^+\rightarrow j j )$&  \\
    \cline{1-1} \cline{3-3}
    $M_{\phi}$=500 GeV  & $ (\phi_1^{-}\rightarrow \Sigma^{0} l^{-}, \Sigma^{0}\rightarrow W^- l^+, W^-\rightarrow l^- \bar{\nu})$   &  0.044\\
     \cline{1-1} \cline{3-3}
    $M_{\phi}$=1000 GeV & $(5l+2j+MET)$  &  0.0014 \\\hline 
    $\rm S2:$& $pp \rightarrow \phi^{++} \phi^{-}_{2} $  &  0.235\\
    $M_{\phi}$=500 GeV & $( \phi^{++}\rightarrow \Sigma^{++} \nu, \Sigma^{++}\rightarrow W^+ l^+, W^+\rightarrow l^+ \nu)$&  \\ \cline{1-1} \cline{3-3}
     $M_{\phi}$=800 GeV & $(\phi^{-}_{2} \rightarrow  \Sigma^{--} l^{+}, \Sigma^{--}\rightarrow  W^- l^- , W^- \rightarrow\, j j) $& 0.022  \\ \cline{1-1} \cline{3-3}
    $M_{\phi}$=1000 GeV& $(4l+2j+MET)$&  0.006  \\
    \hline
& Backgrounds &$\sigma\times$ BR (fb) \\
    \hline
     $\rm B1$& $p p \rightarrow t\bar{t}Z;\; Z \rightarrow l^+ l^-;\; (t \rightarrow b W^+;\;W^+ \rightarrow l^+ \nu;);$& 2.39\\ 
     & $(\bar{t} \rightarrow \bar{b} W^-;\;W ^-\rightarrow l^- \bar{\nu};); $ & \\ \hline
    $\rm B2$& $p p \rightarrow tW^-Zj;\;Z \rightarrow l^+ l^-;\;W^- \rightarrow l^- \bar{\nu};   $  &0.477\\
    & $(t \rightarrow b W^+;\;W^+ \rightarrow l^+ \nu;)$& \\ \hline
     $\rm B3$& $p p \rightarrow ZW^-W^+jj;\; Z \rightarrow l^+l^-;\;W^- \rightarrow l^- \bar{\nu} $  & 0.299 \\ \hline
      
        $\rm B4$ & $p p \rightarrow ZZW^-jj;\; Z \rightarrow l^+l^-;\;W^- \rightarrow l^- \bar{\nu} $ &0.032 \\ \hline
        $\rm B5$& $p p \rightarrow t\bar{t}W^-W^+;\;(t \rightarrow b W^+;\;W^+ \rightarrow l^+ \nu;);\;W^- \rightarrow l^- \bar{\nu} $  & 0.024\\ \hline
         $\rm B6$ & $p p \rightarrow t\bar{t}t\bar{t};\; (t \rightarrow b W^+;\;W^+ \rightarrow l^+ \nu;) $ & 0.021 \\ \hline
     \hline
    \end{tabular}
    \caption{The cross sections $(\sigma \times$ BR) for the two signal (S1 and S2) processes are presented for three benchmark points. Background processes are listed as (B1-B6) and their cross sections $(\sigma \times$ BR) are also given.}
    \label{tab:xsec}
\end{table} 

%---------------------------------------
\iffalse
\begin{comment}

We focused on the pair production of $\phi_1$ as our first signal process(S1), which further decays to 5 leptons and two jets final state,  through the production of $\Sigma^{0}$ from $\phi_1$ at the intermediate. Additionally,  we also took into consideration the production of the $\phi_2$ along with $\phi^{++}$($\phi^{--}$) leading to a final state with four leptons and two jets as a second signal(S2). For S2 the scaler decays to W-boson through the production of $\Sigma^{++}$(and its charge conjugate) at the intermediate state. Table \ref{tab:xsec} lists all the signals and the backgrounds along with their decay channels considered in this work.

We generated these processes with {\tt MadGraph} (version 3.4.1)\cite{Alwall:2014hca} for the center of mass energy 14TeV. {\tt PYTHIA8} (version 8.310) \cite{Sjostrand:2007gs} is used for parton showering, hadronisation, Initial State Radiation (ISR) and Final State Radiation (FSR). The collider effect was added to the events using the CMS card in {\tt Delphes} (version 3.5.0) \cite{deFavereau:2013fsa}. Jets were reconstructed with {\tt FastJet} anti-KT algorithm, with radius parameter 0.5 and $p^{min}_{T,j}$ = 20 GeV. 
    
\end{comment}
\fi
%------------------------------------ 
\paragraph{Detector Simulation:}
The event samples for both signals and backgrounds are generated (in leading-order in $\alpha_s$) using {\tt MadGraph} (version 3.4.1)\cite{Alwall:2011uj} at a center-of-mass energy of 14~TeV. The effects of parton showering, hadronization, as well as initial state radiation (ISR) and final state radiation (FSR) are simulated using {\tt PYTHIA8} (version 8.310)\cite{Bierlich:2022pfr}. The detector-level effects are incorporated using {\tt Delphes} (version 3.5.0)\cite{deFavereau:2013fsa}, employing the default CMS detector configuration. The jets are reconstructed using the anti-$k_{T}$ algorithm as implemented in {\tt FastJet} \cite{Cacciari:2005hq,Cacciari:2011ma} with a radius parameter $R = 0.5$ and a minimum transverse momentum for the jets set at $p^{\text{min}}_{T,j} = 20~\text{GeV}$. All events are required to pass through a few basic selection cuts for the leptons and jets, as: $p_{T,l}>10$ GeV, $|\eta_{l}| < 2.5$ and $p_{T,j}>30$ GeV, $|\eta_{j}| < 4.5$, where $p_{T,l}$ , $p_{T,j}$ are the transverse momentum of the leptons and jets, and $\eta_{l}$, $\eta_{j}$ are their pseudorapidities, respectively. The jets are tagged as b-jets using the default b-tagging module of {\tt Delphes} CMS card.

\paragraph{Pre-Selection : }

In our analysis, leptons play an important role and therefore, a proper identification and selection strategy for the leptons is extremely crucial. Given the mass difference between the scalar and the fermion, the use of the MET direction in identifying the leptons turned out to be an efficient methodology. We, therefore, arrange the leptons in an increasing order of magnitude of $\Delta \phi _{l,MET}$, considering that the lepton with the smallest separation in  $\Delta \phi _{l,MET}$ is produced by the decay of W-boson. To further iterate this point, let us consider the process S2, which has two sources of missing energy; one from the decay of $\phi^{++}$ and another from W-boson decay. 
However, the mass difference between $\phi^{++} \; \rm and \; \Sigma^{++}$ is smaller than that of $\Sigma^{++} $ and the W-boson. Hence, it is expected that the lepton from the decay of $W$ will be identified in the event as the $0^{th}$ lepton ($l_0$), following the above mentioned rule of numbering the leptons. Hence, we obtain 4 leptons as ($l_0$,$l_1$,$l_2$,$l_3$, $l_4$), where $l_0$ is the closest to the MET direction, or in other words, farthest from the di-jet branch, where W decays to jets.

\begin{figure}[!h]
    \centering
    \includegraphics[width=5.5cm,height=5cm]{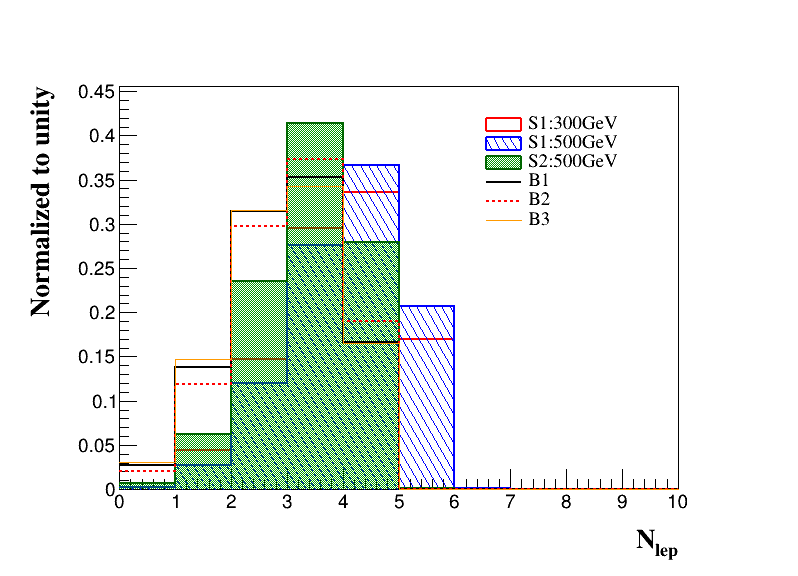}
     \includegraphics[width=5.5cm,height=5cm]{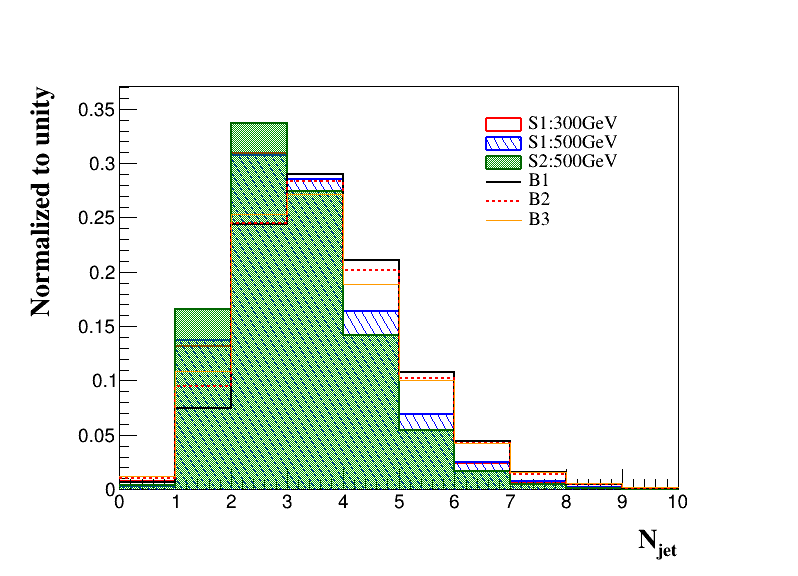}
      \includegraphics[width=5.5cm,height=5cm]{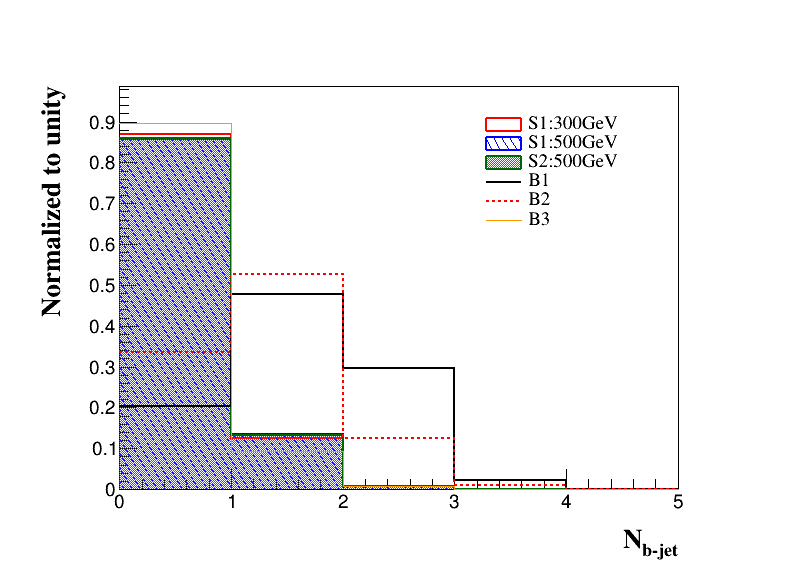}
       \includegraphics[width=6.2cm,height=5cm]{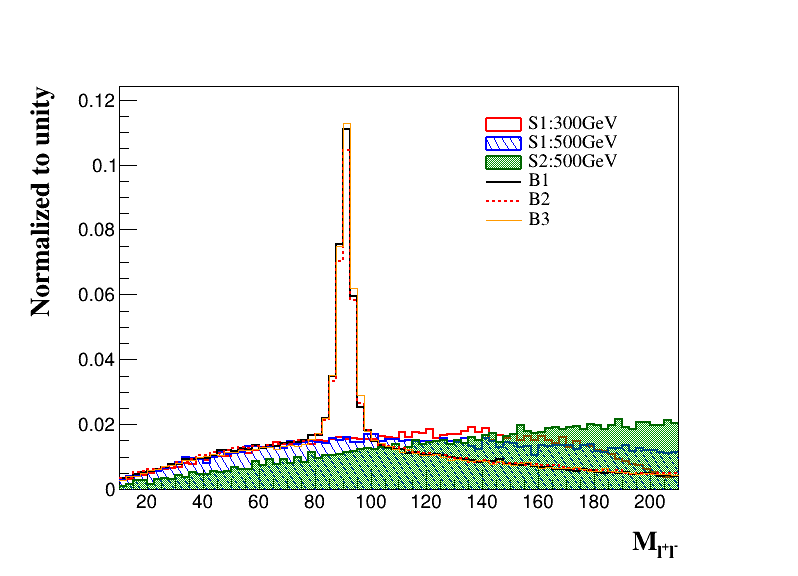}
        \includegraphics[width=6.2cm,height=5cm]{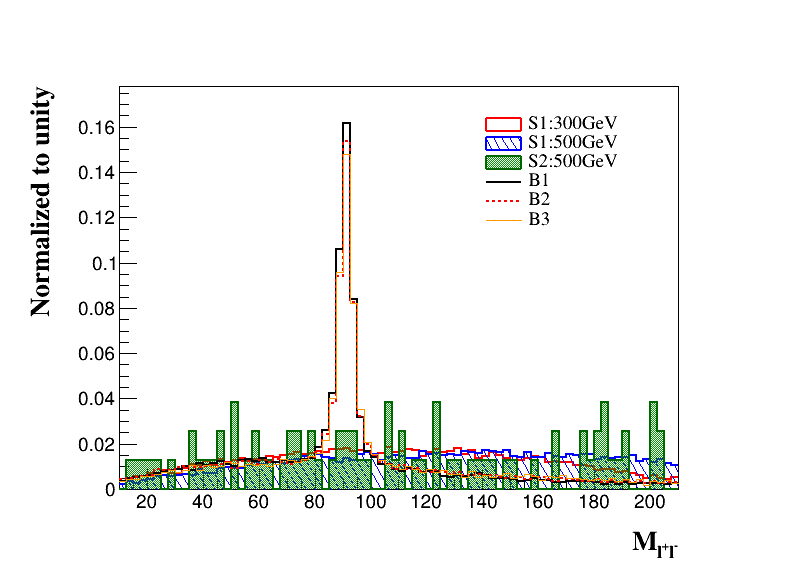}
    \caption{Normalized distributions of the variables used in the pre-selection:  Lepton multiplicity (Top left), jet multiplicity (Top middle),  b-jet multiplicity (Top right ), the invariant mass of the two opposite-sign lepton pairs (Bottom left and Bottom right).
    }
    \label{fig:selection-cuts}
\end{figure}

In the upper panel of Fig.~\ref{fig:selection-cuts}, we show the normalized distribution of the multiplicity of leptons ($N_{lep}$), jets ($N_{jet}$), and b-tagged jets ($N_{b-jet}$). The signal S1 is plotted for two reference points of scalar mass viz., 300 and 500 GeV, while S2 is plotted for the mass 500 GeV. For the backgrounds, only the first three dominant ones are considered in these plots, which helps to identify the appropriate cuts to be applied for event selection and in the optimization process. Several key points to note from these three distributions: \\
(i) The lepton multiplicity distribution peaks at four for S1 and at three for S2, even though these events are generated with five and four leptons, respectively, at the truth level. The answer lies in the distribution of jet multiplicity, where we observe that almost 60\% of the events have a jet multiplicity greater than two for both signals, indicating a smaller number of isolated leptons than expected. \\
(ii) The b-jet multiplicity is nonzero in processes due to the b-tagged jets produced from the decay of the top quark as well as from the decay of the W bosons. The b-jet multiplicity is shown in
Fig.~\ref{fig:selection-cuts} for dominating backgrounds B1, B2, B3 along with the signals.\\
(iii) As discussed before, the invariant mass of the leptons could play an important role in isolating signal events from the background ones. Therefore, we plot the invariant mass of the leading (ordered with respect to the direction with MET) two opposite-sign lepton pairs ($M_{l^+ l^-}$), in the lower panel of  Fig.~\ref{fig:selection-cuts}. 
Most of the backgrounds, B1, B2, B3 and B4, involve a Z-boson which can be reconstructed using opposite-sign leptons.
Hence, we observe a clear resonance peak at 90 GeV for the backgrounds, unlike the signal events, where the leptons are not coming from the decay of the $Z$ boson.
\\
After analyzing these distributions in Fig.~\ref{fig:selection-cuts}, we incorporate a series of cuts, labelled as Pre-selections, to improve the discrimination power of the classifier in discriminating the signal events from the backgrounds, as mentioned in Table \ref{tab:preselection}. 

%------------------------
\begin{table}[!htb]
\centering
\begin{tabular}{|c|c|}\hline 
Pre-selection & $N_{lep}$ $\ge$ 4,  ~$N_{jet}$ $\ge$ 2, ~ $N_{b-jet} =$ 0, ~ $M_{\ell^+\ell^-} \ne [85,95]~{\rm GeV}$\\ \hline
    \end{tabular}
\caption{The list of cuts used as Pre-selection.}
\label{tab:preselection}
\end{table}

%-----
%%%%%%%%%%%%%%%%%%%%%%%%%%%%%

\paragraph{Kinematic Features :}
After applying the selection cuts mentioned above, we proceed to analyze several kinematic features that play an important role in understanding the signal and background topologies. The key observables considered in this study are listed in Table \ref{tab:inputs}. Some of these variables are plotted for the signals and the leading backgrounds in Fig.~\ref{fig:pT} and Fig.~\ref{fig:mass}. In what follows, we discuss some of these features briefly. 

%--------------------------------------
\begin{table}[!htb]
\centering
\begin{tabular}{lllll}   
\toprule
\multicolumn{5}{l}{\textbf{Features used for  BDT }} \\  
\midrule
$\rm p_{T}^{\rm \ell_0}$ & $\rm p_{T}^{\rm \ell_1}$ & $\rm p_{T}^{\rm \ell_2}$ &  $\rm p_{T}^{\rm \ell_3}$ &  $\rm M_{\ell\ell}$ \\ $\rm \cancel{P_{T}}$ & $\rm M_{jj}$ & $\rm \Delta R_{jj}$ & $\rm M_{jj\ell}$ & $\rm M_{jj\ell\ell}$ \\ 
\bottomrule  
\end{tabular}
\caption{\label{tab:inputs}{The set of input variables used for the  multivariate analysis. }}
\end{table}

\begin{itemize}

\begin{figure}[!h]
    \centering
    \includegraphics[width=5.5cm,height=4.8cm]{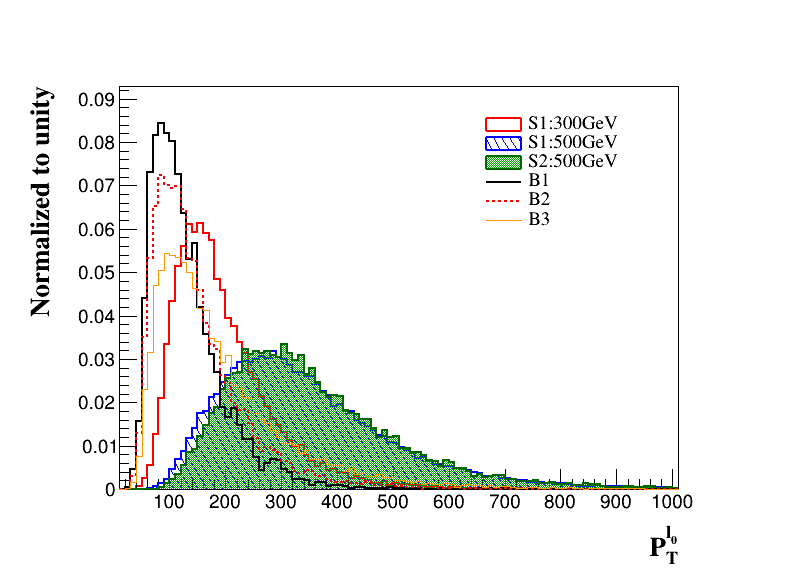}
     \includegraphics[width=5.5cm,height=4.8cm]{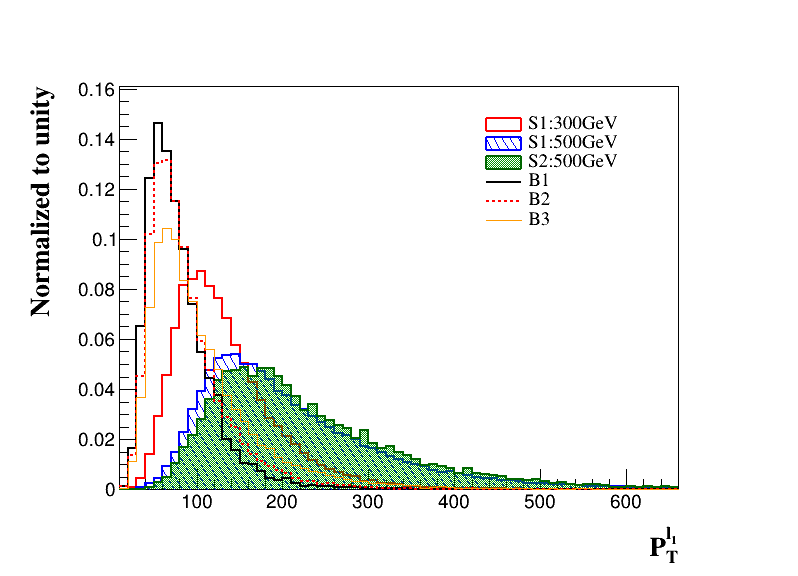}
      \includegraphics[width=5.5cm,height=4.8cm]{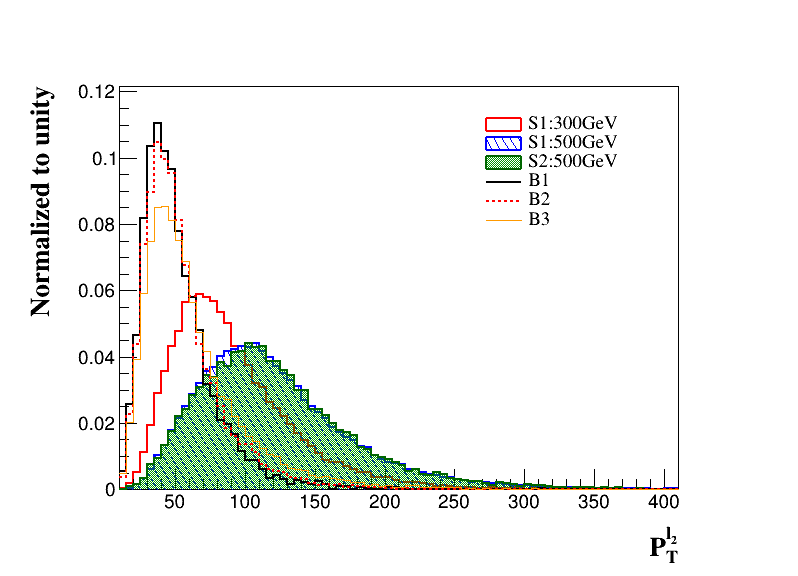}
     \includegraphics[width=5.5cm,height=4.8cm]{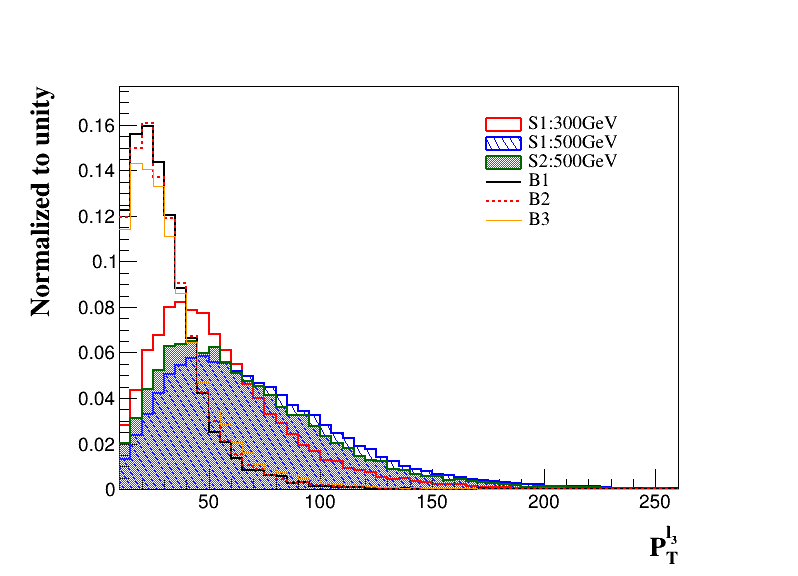} 
     \includegraphics[width=5.5cm,height=4.8cm]{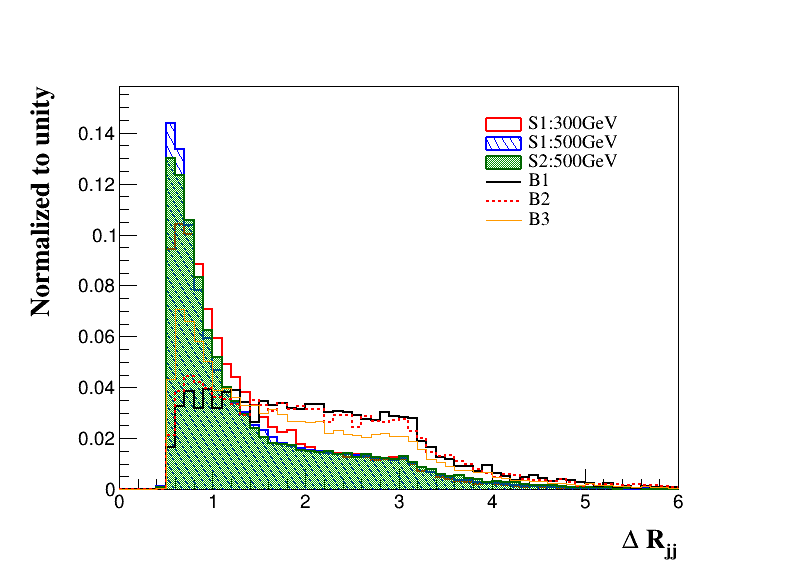}
      \includegraphics[width=5.5cm,height=4.8cm]{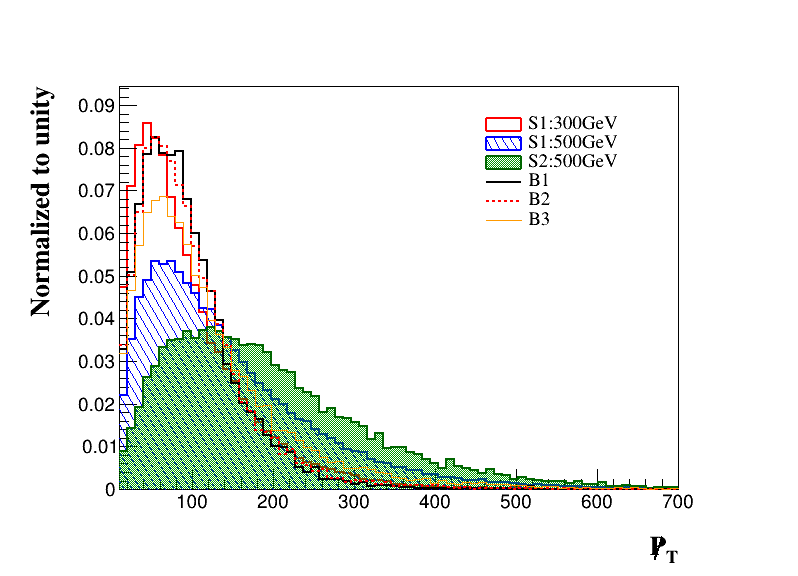}
    \caption{Normalised distributions for Signal and Backgrounds:The angular separation between two jets($\Delta R_{jj}$) (bottom middle), Missing Transverse momentum($\cancel{p_T}$) (bottom right ) and Transverse momentum of four leptons in decreasing order of $p_T$ (top left, top middle, top right and bottom left panel respectively).}
    \label{fig:pT}
\end{figure}  
    \item \textbf{Angular separation between two jets ($\Delta R_{jj}$):} 
  
    Signal events (S1 and S2) produce jets from the decay of a W boson that originated from a heavy fermion, and therefore, these jets are expected to be collimated. As expected, we observe a peak around $\Delta R_{jj}=0.5$ for signal events with a sharply falling distribution after the peak (see the lower middle figure of Fig.~\ref{fig:pT}). The jets from the background events have a relatively flat distribution in $\Delta R_{jj}$. This characteristic difference in the distribution of $\Delta R_{jj}$ makes it an important variable to distinguish signal events from background events.  

    \item \textbf{Invariant mass of a pair of jets ($M_{jj}$):} Following the argument given above, the invariant mass of the leading pair of jets (having the smallest $\Delta R$ between them) is also expected to peak at the W-boson mass. The peak is clearly visible for the signal events in the distribution shown in the top right panel of Fig.~\ref{fig:mass}, unlike the background events. 
    
 \item \textbf{$p_{T}$ of the leptons:}
  As discussed before,  we identify the leptons as $l_0$, $l_1$, $l_2$, $l_3$, and $l_4$, where $l_0$ is the closest to the MET direction, or in other words, farthest from the di-jet branch, where W decays to jets. We further arrange the leptons in the increasing value of $p_{T}$ and we show the distributions in Figure~\ref{fig:pT}. However, we do not use a different notation, we keep the previous notation for the ordered leptons. 
  
\item \textbf{Invariant mass of a pair of leptons ($M_{ll}$):} 
The same-sign leptons do not originate from any particular decay chain. In the signal processes, same-sign leptons can arise from the decays of exotic particles or W, whereas for the backgrounds, same-sign leptons originate predominantly from Standard Model $W$ and $Z$ boson decays. Figure~\ref{fig:mass} (top right panel) shows the invariant mass distribution of the two leading same-sign leptons (ordered according to their direction with respect to the missing transverse energy (MET)), for both the signal and background processes. The pattern of the invariant mass distributions exhibits a significant distinction between signal and background topologies.
As the scalar mass increases, the signal distribution shifts towards higher masses relative to the background. Furthermore, for the same benchmark mass of 500 GeV, the distribution for S2 is more separated from the background distributions compared to S1.

    \item \textbf{Invariant mass of a pair of jets with lepton(s):} The presence of two exotic particles in the signal events hints at the different patterns for the energy flow among the different final state particles, especially the leptons and jets. Therefore, we calculate two invariant mass observables, the first involving a pair of jets and a lepton ($M_{jj\ell}$), and the other involving a pair of jets with a pair of leptons ($M_{jj\ell\ell}$). 
    
    To demonstrate the reconstruction process, let us take the example of S1. To calculate these, we consider a pair of leptons that are closer to the jet system and further from the MET direction. Note that, for signal events, the two closest leptons to the jet pair are the decay products of $\phi_1$ and $\Sigma^0$ respectively.
    Furthermore, among the two leptons, the lepton from the decay of $\Sigma^0$ possesses larger $p_T$ than the one of $\phi_1$-decay, as the mass difference between $\Sigma^0$ and the W-boson is much larger than the mass difference between $\phi_1$ and $\Sigma^0$. 
    The $M_{jj\ell}$ distribution is shown in the bottom left panel of Fig.~\ref{fig:mass} displays a sharp peak at the representative mass of the $\Sigma^0$-particle. The mass of $\Sigma^0$ is reconstructed at three different benchmark points. For S2, the reconstruction process of the invariant masses are similar to S1.

     Then we  calculate the invariant mass of the jet pair along with the two leptons closest to the jet pair, namely $M_{jj\ell\ell}$. The signal distribution for $M_{jj\ell\ell}$ shows a sharp peak at the $\phi$ masses, as shown in bottom right panel of Fig.~\ref{fig:mass}. The background events display a wide spread in the said distributions (see the bottom panel of Fig.~\ref{fig:mass}). 

     The rest of the two leptons that do not take part in the reconstruction are tagged.
To maximize the discrimination power, we also include the $p_T^{\ell_i}$, and the missing transverse momentum of the events to the machine learning algorithm, as listed in Table \ref{tab:inputs}. For completeness, we have included the distribution of the MET in Fig.~\ref{fig:pT}, bottom right panel.  
    
\end{itemize}
%-----------------------------------------------
\begin{figure}[!htb]
    \centering
        \includegraphics[width=0.45\linewidth]{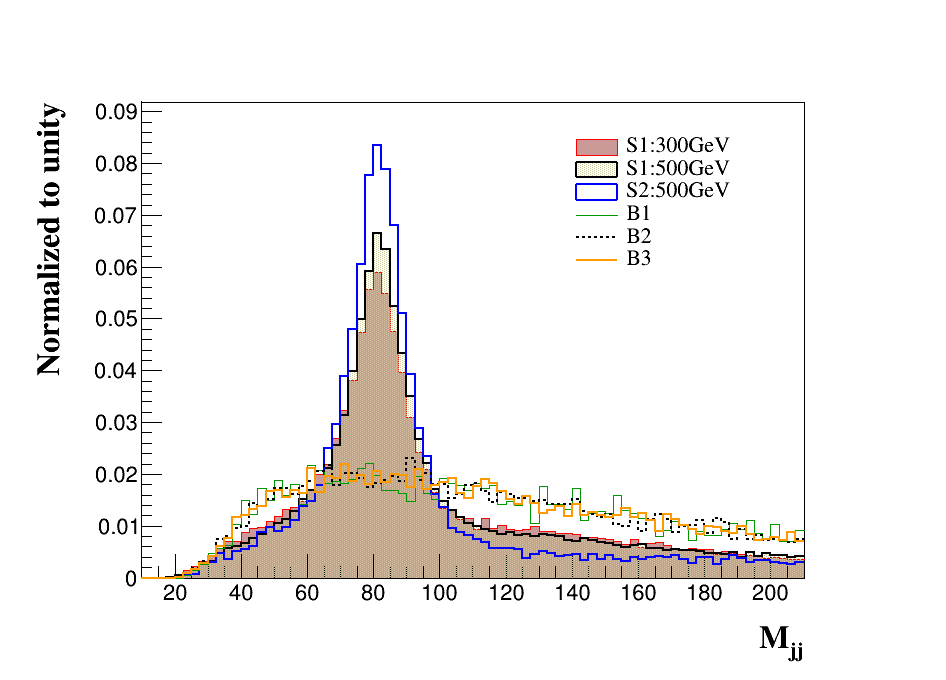}
     \includegraphics[width=0.45\linewidth]{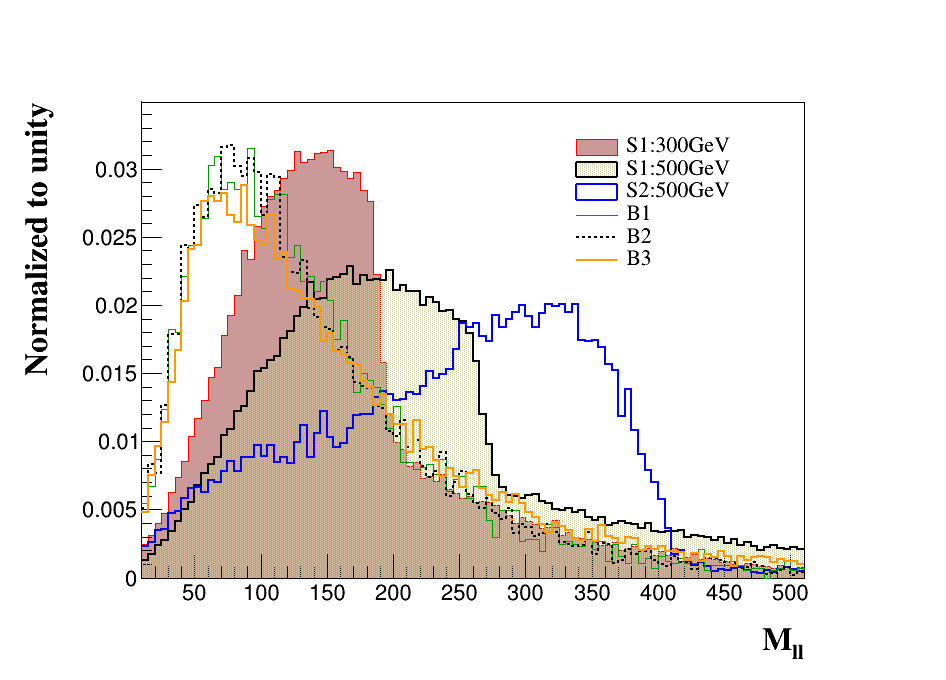}
      \includegraphics[width=0.45\linewidth]{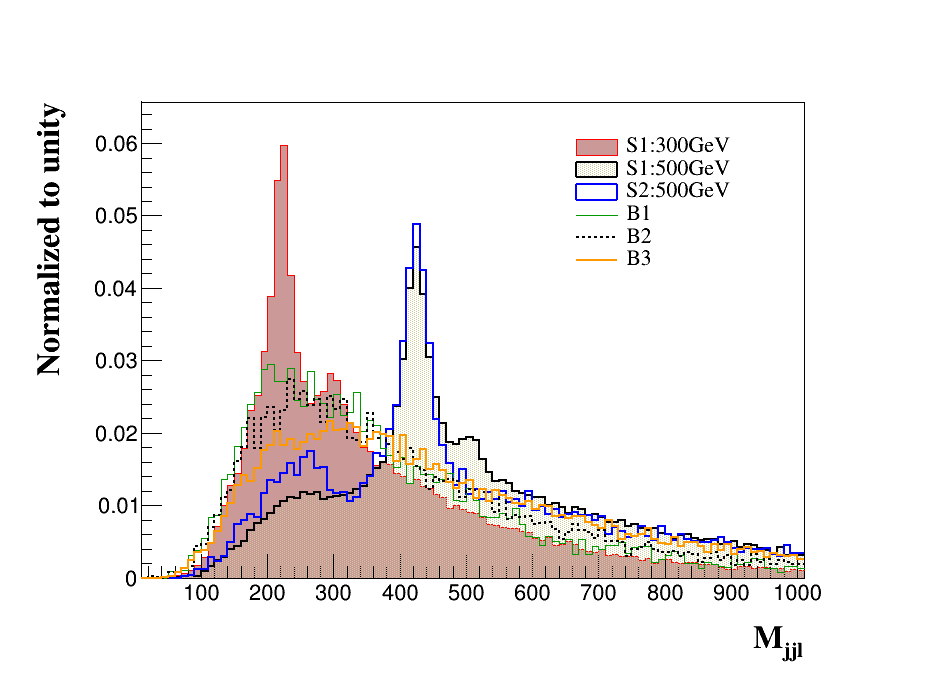}
      \includegraphics[width=0.45\linewidth]{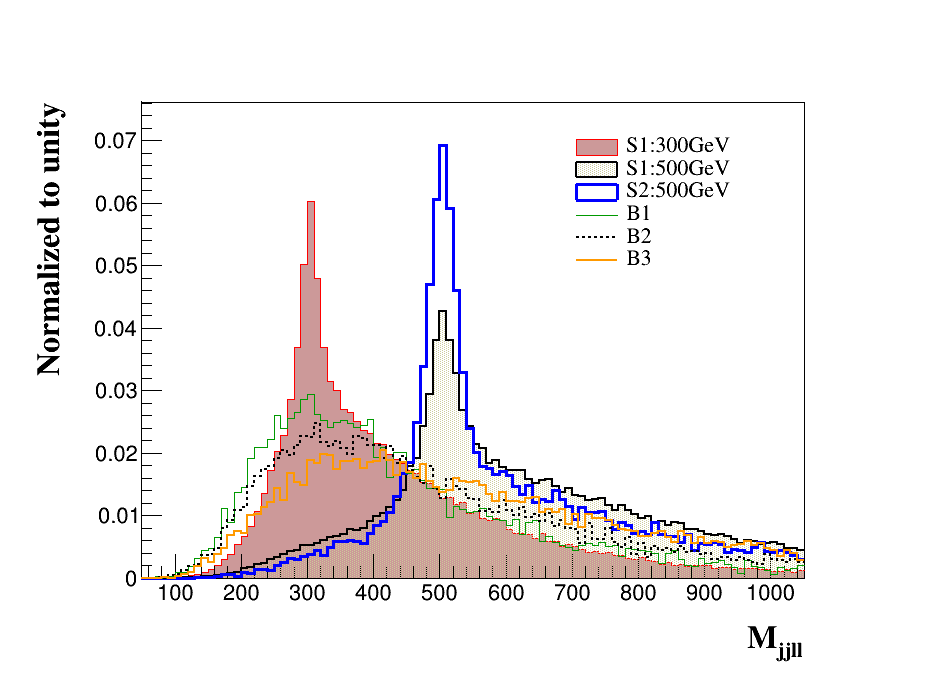}

    \caption{Normalised distribution of invariant mass distributions of lepton-jet combinations for signal and background processes. Invariant mass of two jets (top left), invariant mass of two same sign leptons(top right), three body invariant mass of the jet pair with one lepton (bottom left), four body invariant mass of the jet pair with two leptons (bottom right). 
    }
    \label{fig:mass}
\end{figure}

%%%%%%%%%%%%%%%%
\paragraph{Result:}
Understanding the signals at different benchmark points leads us to identify the features useful for discriminating the signal from the backgrounds having similar final states but higher cross sections. Now, we focus on building a classification model, which can help the signals achieve an impressive significance even if the cross section is small. 
The combined data set of a signal for a specific benchmark point, together with the backgrounds, are divided into a $50:50$ ratio for training and testing in such a way that for training, there are equal numbers of signal and background events present. Each of the background is weighted according to their cross section with the weight = $\frac{1}{N_{MC}} \times \sigma_{MC} \times \mathcal{BR} \times (\int \mathcal{L} dt)$, where $N_{MC}$ is the number of Monte Carlo samples generated, $\sigma_{MC}$ is the production cross section of each process, $\mathcal{BR}$ is the branching ratio for a particular decay channel and $\int \mathcal{L} dt$ is the integrated luminosity. The analysis is carried out on the high luminosity run of LHC with $\int \mathcal{L} dt \;=\; 3000 fb^{-1}$. All features or observables are passed through the MVA, and a Receiver Operating Characteristics (ROC) curve is obtained that gives the probability of correctly rejecting background events as a function of the probability of correctly identifying signal events. We use the ROOT Library, ``Toolkit for Multivariate Analysis'' \cite{TMVA:2007ngy} to
perform the MVA. More specifically, we use the algorithm ``Adaptive Boosted Decision Tree'' (AdaBoost) \cite{FREUND1997119} for the classification of signal and background events.
In Fig.~\ref{fig:ROC}, we show the ROCs obtained for the signal process S1 and S2 for two representative values of the scalar mass. As evident from the figure, the higher the scalar mass, the better the separation in the input observables, leading to improved performances in terms of the Area Under the Curve (AUC). 

To estimate the discovery potential of the HL-LHC, we calculate the signal significances for each benchmark point, defined as (also referred to as the Asimov significance \cite{Cowan:2010js}):$$\mathcal{S} = \sqrt{2 \left[ (S + B) \ln\left(1 + \frac{S}{B}\right) - S \right]},$$
 with S and B being the number of signal and background events, respectively, after the imposition of the BDT cut. Essentially, it quantifies the signal's distinguishability from expected background noise. The  significance of the signal S1 and S2 are listed in Table \ref{Tab:signi}. 
 We observe that the discovery prospect (i.e., $\mathcal{S}\geq 5$) for the scalars through the process S1 lies up to approximately 600 to 700 GeV, while for exclusion $\mathcal{S}\gtrsim 2$ level, the limit can be extended up to $\sim 900$ GeV. We have found that the significance is larger in S2 compared to S1, even though the cross section is smaller in S2.
 Using S2, we can exclude ($\mathcal{S}\gtrsim 2$) a new scalar up to a mass of approximately 1 TeV (1000 GeV) or more.
%---------------------------------
 \begin{figure}[!htb]
    \centering
    \includegraphics[width=0.45\linewidth]{figures_pheno/roc_4l2j_04.png}
 \caption{Receiver Operating Characteristic (ROC) curves displaying the signal efficiency ($\epsilon_s$) on the x-axis and the background rejection ($1-\epsilon_b$) on the y-axis for the two signal processes S1 and S2 at different benchmark points. }
    \label{fig:ROC}
\end{figure}
%----------------------------------
\begin{table}[!htb]
\centering
\begin{tabular}{|c|c|c|c|}
\hline
Process & $M_{\Phi}$ & BDT Cut & \(\mathcal{S} \,\) \\ 
& (GeV) &    & \\\hline
S1&300&0.1&33.7 \\  \cline{2-4}
&500&0.1 & 9.17 \\ \cline{2-4}
&1000&0.2 & 0.88\\ \hline\hline
S2&500&0.1 &  19.7\\ \cline{2-4}
&800& 0.2 &  4.8 \\ \cline{2-4}
&1000& 0.2 &  2.1 \\ \hline
\end{tabular}
\caption{Signal significances with the selection cuts mentioned in Table \ref{tab:preselection} for integrated luminosity 3000 ${\rm fb}^{-1}$ for at different benchmark points in S1 and S2 along with the respective BDT cuts for which the significance is maximum.
}
\label{Tab:signi}
\end{table}
%====================================

We would also like to mention that the lepton multiplicity distribution ($N_{lep}$), as shown in Figure \ref{fig:selection-cuts} indicates that a requirement $N_{lep} \ge 5$ effectively suppresses the SM background processes. We have checked that a similar analysis carried in this channel suppress B1, B2, and B3 to negligible levels. The major contribution in the background comes from B4, which has cross section similar to $S1$ at 500 GeV mass. as the primary remaining contamination. Therefore, we find that the cut significantly reduce background contamination, leading to a noticeable improvement in signal significance in both S1 and S2. We find that with this revised search strategy, the discovery prospect for S1 reaches up to $\sim 600-700$ GeV. while the exclusion limit reaches $\sim 900$ GeV. However, the signal cross section is also significantly reduced compared to the previous analysis. With so few events, reconstruction of either the fermion or the scalar states is not feasible. Therefore, we do not pursue this channel in further detail.

%=====================================================
\section{Summary} 

In this work, we consider a simple extension of the SM with a scalar quadruplet and a fermionic quintuplet, motivated from the resolution to neutrino mass problem in the SM. In the collider study, we focus on the production and decays of the quartet scalar. The decay pattern of the quartet components depends upon the mass hierarchy between the quartet scalar and the quintuplet fermion as well as the Yukawa coupling between the two.  We consider the case when $M_\Phi > M_\Sigma$. We observe two different decay patterns wherein the scalar can either decay to gauge bosons directly or they first decay to quintuplets which subsequently decay to gauge bosons. For large Yukawa coupling values ($\sim 1$), scalars predominantly decay to quintuplets; this regime is referred to as {\it fermiophilic}. As Yukawa coupling strength decreases, scalar decays to gauge bosons become increasingly significant. In the limit of very small $y_\nu$ ($\leq 0.0001$), scalars decay predominantly to gauge bosons, defining the {\it fermiophobic} scenario.

We explore the prospective collider signatures in these two scenarios, alongside the prevailing experimental constraints from existing collider searches. We find that the signals with 4 or more leptons along with 2 jets offer substantial signal $\times$ BR and are not excluded by the LHC experiment. 
We study the pair production mode $\phi_1^+\phi_1^-$ and associated production mode $\phi^{++}\phi_2^-$ and their decays via the fermion quintuplets at $14$ TeV LHC, giving rise to final states with  4 or more leptons along with 2 jets. We focus on the {\em fermiophilic} scenario and demonstrate that scalar and fermion masses can be reconstructed through a selection strategy based on kinematic variables, applied to the final-state particles.

 We also conduct a detailed collider analysis for the final state with at least four leptons and two jets, incorporating SM background events within a multivariate analysis (MVA) framework, to assess the discovery prospects of these scalars. The pair production process for the scalars achieves a sensitivity of $5\sigma$ with mass of the scalar ranging from $\approx 600$ to $700\text{ GeV}$ and an exclusion limit ($2\sigma$) up to $\approx 900\text{ GeV}$. Critically, the associated production process offers superior sensitivity, extending the exclusion potential ($2\sigma$) for the scalar mass to $1\text{TeV}$ and beyond. The signal processes feature cascade decays that pose significant challenges to reconstruction. We have also demonstrated that the successful reconstruction of quadruplet scalars and quintuplet fermions is feasible in both channels, S1 and S2. 

 We also examine the discovery and exclusion reach in the $5\ell+2$ jets final state. Although this channel yields even higher significance, the number of surviving signal events after all selections falls below 10 at $\mathcal{L} = 3000$ fb$^{-1}$. Furthermore, for scalar masses exceeding 1 TeV, the signal cross section decreases substantially, rendering discovery unattainable in these final states. Also, at higher energies, the $W$ bosons become highly energetic and could be reconstructed as fat jets; however, such an analysis lies beyond the scope of this work.
\section{Acknowledgments}
%=============================================
N.K. would like to thank Anusandhan National Research Foundation
Government of India for the support from Core Research Grant with grant agreement number CRG/2022/004120. 
\bibliographystyle{unsrt}
\bibliography{ref}
%=============================

\end{document}